\documentclass[a4paper,aps,prd,twocolumn,10pt,superscriptaddress,nofootinbib,showpacs]{revtex4-1}
 \pdfoutput=1
\usepackage{graphicx,amsmath,amsfonts,amssymb,revsymb,dcolumn,epsfig,bm,xspace}
\usepackage[hyperfootnotes=false]{hyperref}
\usepackage[latin1]{inputenc}
\usepackage{color}
\usepackage{bbold}

\newcommand{\OO}[1]{\mathcal{O}\left(#1\right)}
\newcommand{\dd}{\mathrm{d}}
\newcommand{\ex}{\mathrm{e}}
\newcommand{\lie}{\pounds}
\newcommand{\del}{\partial}
\newcommand{\dirac}[2]{\delta^{(#1)}\!\left(#2\right)}

\newcommand{\Comm}[2]{\left[#1,\;#2\right]}

\newcommand{\indspace}{\quad}
\newcommand{\Prod}[2]{\left(#1, #2\right)}
\newcommand{\ci}{\mathsf{i}}

\DeclareFontFamily{U}{mathx}{\hyphenchar\font45}
\DeclareFontShape{U}{mathx}{m}{n}{<-> mathx10}{}
\DeclareSymbolFont{mathx}{U}{mathx}{m}{n}
\DeclareMathAccent{\widebar}{0}{mathx}{"73}
\newcommand{\backdec}[1]{\bar{#1}}
\newcommand{\wbackdec}[1]{\widebar{#1}}
\newcommand{\g}{g}

\newcommand{\bg}{{\backdec{g}}}
\newcommand{\n}{u}
\newcommand{\bn}{{\backdec{\n}}}

\newcommand{\h}{\gamma}
\newcommand{\bh}{\backdec{\h}}

\newcommand{\hp}[1]{\h\left[#1\right]}
\newcommand{\bhp}[1]{\bh\left[#1\right]}
\newcommand{\cd}{\nabla}
\newcommand{\bcd}{\wbackdec{\nabla}}
\newcommand{\scd}{D}
\newcommand{\bscd}{\wbackdec{\scd}}
\newcommand{\scp}{\parallel}
\newcommand{\blap}{\wbackdec{\scd}^2}
\newcommand{\bclap}{\hat{\scd}^2}
\newcommand{\bclapK}{\hat{\scd}^2_{\K}}
\newcommand{\blapK}{\blap_{\K}}
\newcommand{\bilap}{\wbackdec{\scd}^{-2}}
\newcommand{\bilapK}{\bilap_{\K}}
\newcommand{\blapilapK}{\Delta_\K}

\newcommand{\R}{R}

\newcommand{\bR}{\wbackdec{\R}}
\newcommand{\G}{G}
\newcommand{\bG}{\wbackdec{\G}}
\newcommand{\kp}{\kappa}

\newcommand{\ddg}{\xi}

\newcommand{\LL}{{\mathcal F}}

\newcommand{\A}{\phi}
\newcommand{\B}{B}
\newcommand{\C}{C}
\newcommand{\CS}{\psi}
\newcommand{\CSI}{\Psi}
\newcommand{\CSvms}{\mathcal{U}}
\newcommand{\CSvmsi}[1]{\CSvms_{#1}}
\newcommand{\CSvmsidt}[1]{\dot{\CSvms}_{#1}}
\newcommand{\CSvmsmi}[1]{\widetilde{\CSvms}_{#1}}
\newcommand{\CSvmsmidt}[1]{\dot{\widetilde{\CSvms}}_{#1}}
\newcommand{\MS}{\zeta}
\newcommand{\CST}{\CS^{\mathsf{t}}}
\newcommand{\CTD}{W}
\newcommand{\BS}{\mathcal{\B}}
\newcommand{\BV}{\mathtt{\B}}
\newcommand{\CSD}{\mathcal{E}}
\newcommand{\CV}{\mathtt{F}}

\newcommand{\EC}{\mathcal{K}}
\newcommand{\bEC}{\wbackdec{\EC}}

\newcommand{\EX}{\Theta}
\newcommand{\dEX}{{\delta\EX}}
\newcommand{\dEXi}{\Xi}

\newcommand{\bEX}{\wbackdec{\EX}}

\newcommand{\dSH}{\delta\sigma}
\newcommand{\dSHs}{\dSH^{\text{s}}{}}
\newcommand{\dSHv}{\dSH^{\text{v}}{}}

\newcommand{\SR}{\mathcal{R}}
\newcommand{\dSR}{{\delta\SR}}
\newcommand{\bSR}{\wbackdec{\SR}}
\newcommand{\K}{K}
\newcommand{\bK}{\wbackdec{\K}}

\newcommand{\gH}{\mathcal{H}}

\newcommand{\mf}{\varphi}

\newcommand{\pmf}{\Pi_\mf}
\newcommand{\phv}{\chi}
\newcommand{\Ophv}{\hat{\chi}}
\newcommand{\OFphv}{\tilde{\chi}}
\newcommand{\Omf}{\hat{\mf}}
\newcommand{\Opmf}{\hat{\Pi}_{\mf}}
\newcommand{\nm}{u}

\newcommand{\vms}{\mathcal{V}}
\newcommand{\vmsi}[1]{\vms_{#1}}
\newcommand{\vmsidt}[1]{\dot{\vms}_{#1}}

\newcommand{\ST}{\Sigma}

\newcommand{\en}{s}

\newcommand{\ed}{\rho}

\newcommand{\bcs}{\backdec{c}_\text{s}}
\newcommand{\bcm}{\backdec{c}_\text{m}}
\newcommand{\bcn}{\backdec{c}_\text{n}}
\newcommand{\bcsi}[1]{\backdec{c}_{\mathrm{s}#1}}

\newcommand{\ded}{\delta\ed}
\newcommand{\dedgi}{\delta\ed^{\text{gi}}}
\newcommand{\bedpf}{\varpi}
\newcommand{\dedc}{\delta_\CS}
\newcommand{\dedci}[1]{\delta_{\CS{}#1}}
\newcommand{\dedcmi}[1]{\tilde{\delta}_{\CS{}#1}}
\newcommand{\dedcmidt}[1]{\dot{\tilde{\delta}}_{\CS{}#1}}

\newcommand{\bp}{\backdec{p}}
\newcommand{\bpi}[1]{\bp_{#1}}
\newcommand{\bed}{\backdec{\ed}}
\newcommand{\bedi}[1]{\bed_{#1}}

\newcommand{\bedpi}[1]{(\bedi{#1}+\bpi{#1})}
\newcommand{\bedp}{(\bed + \bp)}

\newcommand{\LA}{\mathcal{L}}

\newcommand{\dLA}{{\delta\LA}}

\newcommand{\mat}{\text{m}}
\newcommand{\tbg}{\text{bg}}
\newcommand{\HA}{\mathcal{H}}
\newcommand{\dHA}{{\delta\HA}}
\newcommand{\pCSD}{\Pi_{\CSD}}
\newcommand{\pDCSD}{\Pi_{\CSD}}
\newcommand{\pCST}{\Pi_{\CST}}

\newcommand{\pvmsi}[1]{\Pi_{\vms{}#1}}

\newcommand{\pMS}{\Pi_\MS}

\newcommand{\pCSvms}{\Pi_{\CSvms}}
\newcommand{\pCSvmsi}[1]{\Pi_{\CSvms{}#1}}
\newcommand{\pCSvmsmi}[1]{\widetilde{\Pi}_{\CSvms{}#1}}

\newcommand{\HF}[1]{\mathcal{Y}_{#1}}
\newcommand{\EV}[1]{\lambda_{#1}}

\newcommand{\HFi}{\bm{k}}
\newcommand{\BT}{\mathsf{T}}
\newcommand{\BTi}{\mathsf{t}}

\newcommand{\BF}[1]{\mathsf{U}_{#1}}

\newcommand{\AO}{\mathsf{a}}

\newcommand{\ket}[1]{\left\vert{#1}\right\rangle}

\newcommand{\braketOP}[3]{\left\langle#1\middle\vert#2\middle\vert#3\right\rangle}

\newcommand{\ofreq}{\nu}

\newcommand{\om}{m}
\newcommand{\SM}{\mathbb{S}}

\newcommand{\lapse}{\mathcal{N}}
\newcommand{\POT}{\mathsf{V}}
\newcommand{\KIN}{\mathsf{T}}
\newcommand{\iS}{Q}
\newcommand{\ipS}{P_\iS}

\newcommand{\GN}{G_{_\mathrm{N}}}
\newcommand{\MP}{M_{_\mathrm{Planck}}}

\begin{document}
\title{Quantum Cosmological Perturbations of Multiple Fluids}

\author{Patrick Peter} \email{peter@iap.fr}
\affiliation{Institut d'Astrophysique de Paris (${\cal G}\mathbb{R}\varepsilon\mathbb{C}{\cal O}$),
UMR 7095 CNRS,
Sorbonne Universités, UPMC Université Paris 06, Institut Lagrange de Paris, 98 bis boulevard Arago, 75014 Paris, France.}

\author{N.~Pinto-Neto} \email{nelson.pinto@pq.cnpq.br}
\affiliation{Centro Brasileiro de Pesquisas F\'{\i}sicas, Rua
  Dr. Xavier Sigaud 150, Urca,  CEP 22290-180, Rio de Janeiro, Brasil}

\author{Sandro D.~P.~Vitenti} \email{dias@iap.fr}
\affiliation{CAPES Foundation, Ministry of Education of Brazil, Bras\'{\i}lia -- DF 70040-020, Brazil}
\affiliation{Institut d'Astrophysique de Paris (${\cal G}\mathbb{R}\varepsilon\mathbb{C}{\cal O}$),
UMR 7095 CNRS,
Sorbonne Universités, UPMC Univ. Paris 06, Institut Lagrange de Paris,
98 bis boulevard Arago, 75014 Paris, France.}

\date{\today}

\begin{abstract}
The formalism to treat quantization and evolution of cosmological
perturbations of multiple fluids is described. We first construct the
Lagrangian for both the gravitational and matter parts, providing the
necessary relevant variables and momenta leading to the quadratic
Hamiltonian describing linear perturbations. The final Hamiltonian is
obtained without assuming any equations of motions for the background
variables. This general formalism is applied to the special case of two
fluids, having in mind the usual radiation and matter mix which made
most of our current Universe history. Quantization is achieved using an
adiabatic expansion of the basis functions. This allows for an
unambiguous definition of a vacuum state up to the given adiabatic
order. Using this basis, we show that particle creation is well
defined for a suitable choice of vacuum and canonical variables, so that
the time evolution of the corresponding quantum fields is unitary. This
provides constraints for setting initial conditions for an arbitrary
number of fluids and background time evolution. We also show that the
common choice of variables for quantization can lead to an ill-defined
vacuum definition. Our formalism is not restricted to the case where
the coupling between fields is small, but is only required to vary
adiabatically with respect to the ultraviolet modes, thus paving the way
to consistent descriptions of general models not restricted to
single-field (or fluid).
\end{abstract}

\pacs{98.80.Es, 98.80.-k, 98.80.Jk}

\maketitle

\section{Introduction}

Cosmological perturbations are usually thought to originate from the
short wavelength quantum vacuum fluctuations of the matter fields or
fluids and the metric. The scales are subsequently increased to exit the
Hubble radius, whence particles are produced, hopefully resulting in the
observed spectrum of primordial perturbations. In inflationary models,
the usual procedure consists in assuming a
Friedmann-Lema\^{\i}tre-Robertson-Walker (FLRW) background metric with
scale factor evolving following General Relativity (GR) Einstein
equations sourced by a slowly rolling scalar field, leading to a
quasiexponential, almost de Sitter inflation phase. Expanding the
Einstein-Hilbert plus scalar field action to second order in
perturbations, one then easily quantizes the so-called Mukhanov-Sasaki
variable which can, for wavelengths much smaller than the Hubble scale,
be postulated to initiate in a vacuum state: in a de Sitter universe,
the (Bunch-Davies) vacuum state is unambiguously defined. The procedure
can be extended to the quasi-de Sitter case that is of interest for
inflation; this scheme is extremely well suited to the inflationary
setup to which it is regularly applied.

In the alternative bouncing scenarios, primordial perturbations are set
as vacuum initial conditions in the far past of the contracting phase,
when the Universe is assumed almost flat for the scales of interest.
When the contraction is dominated by a pressureless fluid, the spectrum
of long wavelength perturbations is almost scale invariant
\cite{Finelli2002, Peter2007, Peter2008} (and slightly blue). This is
also the case for the so-called ekpyrotic scenarios \cite{Khoury2001,
	Kallosh:2001ai, Khoury2002}. In this case however, the bounce transition
itself can lead essentially to any prediction~\cite{Martin2002}.

Although the currently available data are well explained by single-field
inflation, alternative views exist that should be considered. One might
for instance be interested in investigating multifield inflation,
although Bayesian analysis tend to disfavor them \cite{Martin:2013nzq}.
Indeed, additional fields are quite generic in supposedly realistic
theories such those based on a grand unification or string theory, so
that having more than one field present during the inflationary stage
appears natural. Similarly, realistic bouncing models share the same
characteristic, as one naturally expects several fluids, e.g. radiation
and dust, to be present and active together with that responsible for
the bouncing phase itself. A general rigorous formalism to deal with
such situations is still lacking, although tentative options were
proposed. In the multifield inflation scenario, it has been suggested
to consider the coupling between fields as important only after the
modes have entered the potential which, loosely speaking, is
equivalent to becoming larger than the Hubble radius. It was also assumed that the coupling is present before potential entry, but
is such that there exists a limit where it is asymptotically
negligible, so that the vacuum can be defined as a combination of
decoupled fields. These cases are discussed, e.g., in Refs.~\cite{Byrnes2006,
Lalak2007, Langlois2008, Langlois2008a, Battefeld2011, Ellis2014,
McAllister2012, Mizuno2014, Renaux-Petel2015, Renaux-Petel2015a}. In
both procedures the vacuum is set in the limit where the coupling is
zero, a procedure which cannot take into account a possible divergent
behavior of the couplings in the Ultraviolet (UV) limit. Such a behavior
would lead to a set of ill-defined basis functions and, consequently, an
ill-defined vacuum. To evaluate this behavior, one must calculate the
next-to-leading-order corrections of all basis functions. The purpose of
this work is precisely to calculate these corrections in a general way.

The present work, coming as a follow-up of Refs.~\cite{Vitenti2013,
	Falciano2013, Vitenti2015}, aims at partially closing the gap in
providing this means, at least in cases consisting of many constituents,
assumed uncoupled except through gravity. Our formalism henceforth
permits us to set up initial conditions not only in an inflationary
expanding universe, but also in a contracting universe, taking into
account couplings between dust, radiation, and any other component that
one might want to include in the model, under rather general
circumstances. It is thus particularly suited to bouncing models in
which the Universe starts big and filled with otherwise ordinary fluids.

In the following sections, we first set up the necessary formalism to
describe cosmological perturbations around an FLRW background by
expanding the Lagrangian action up to second order, assuming GR and an
arbitrary number of fluids (Sec.~\ref{sec:L}); our formalism, based on
canonical transformations, does not rely on the background satisfying
Einstein equations, so that even with the Einstein-Hilbert action
describing both background and perturbations, the actual background
equations of motion are not necessarily the classical ones. Performing
the required transformations, we deduce in Sec.~\ref{sec:H} the general
Hamiltonian, then in Sec.~\ref{sec:split} we introduce the
adiabatic/entropy mode splitting, which we illustrate with the special
case of a two-component fluid, assumed barotropic. We find that coupling
through gravity implies a special form of the coupling terms, which is
quadratic in the fluid momenta.

The Hamiltonian is of course the starting point required for
quantization, which we proceed with in Sec.~\ref{sec:Q}. Expanding the
relevant variables in terms of harmonic functions of the
Laplace-Beltrami operator, we derive the appropriate field operators and
their momenta, and impose canonical quantization upon them. Because the
Hamiltonian stems from perturbation theory and is thus quadratic, it
turns out to be convenient to simplify the whole discussion using
symplectic forms and a generic description of the Hamiltonian tensor in
terms of block matrices describing the kinetic and potential terms. In
Sec.~\ref{sec:genapp}, we present the quantization procedure valid for
any system fitting this description and, in Sec.~\ref{sec:vevol}, we
discuss the time evolution for such systems and the canonical
transformations producing a set of variables where the quantization
procedure leads to a unitary evolution of the quantum
fields. Finally, in Sec.~\ref{sec:condini}, we obtain the explicit form
for the adiabatic approximation for the basis functions. This permits
the evaluation of the UV limit order by order in the
adiabatic expansion. Using this tool, we show that, for the usual choice
of variables, the adiabatic limit does not commute with the UV limit and
consequently cannot be used to define a basis for the whole UV spectrum
(which we call UV incomplete). On the other hand, using the variables
introduced in this work, the vacuum is UV complete and the particle
production is finite (in the sense that the particle density is finite).
This result extends the work done in Ref.~\cite{Vitenti2015} (for
particular examples, see also~\cite{Torre2002,Corichi2006, Cortez2007,
	Corichi2007,BarberoG2008,Vergel2008,Cortez2011,
	Gomar2012,Cortez2012,Fernandez-Mendez2012,Cortez2013, Cortez2015}) by
showing a concrete example of choice of variables and initial conditions
which leads to a unitary implementation of the time evolution for
quantum fields interacting through a quadratic term in the Hamiltonian.

\section{Second Order classical theory}

Cosmological perturbations stem from an expansion around a FLRW
background. The classical theory needs be written to second order in the
perturbation variables, and subsequently quantized. In this section we
first present the second order Lagrangian describing the dynamics of
such linear cosmological perturbations of a FLRW universe with matter
content provided by an arbitrary number of noninteracting fluids.

\subsection{Lagrangian formulation}
\label{sec:L}

Classical GR coupled to perfect fluids is most naturally expressed in
terms of an action, deriving from a Lagrangian functional, from which
the Hamiltonian formalism, necessary for quantization, can be derived.
We begin by recalling the basics of the required Lagrangian formalism.

\subsubsection{Conventions}

We consider a spacetime manifold described by a physical metric
$\g_{\mu\nu}$, with signature $(-,+,+,+)$. The torsion-free covariant
derivative compatible with this metric is denoted by $\nabla_\mu$, such
that $\nabla_\mu g_{\alpha\beta} = 0$.\footnote{Our conventions for the
	curvature tensors are:
$$(\nabla_\mu\nabla_\nu - \nabla_\nu\nabla_\mu)v_\alpha =
  R_{\mu\nu\alpha}{}^\beta{}v_\beta, \quad
  R_{\mu\alpha} \equiv R_{\mu\nu\alpha}{}^\nu, \quad
  R \equiv R_\mu{}^\mu,$$ where $v_\alpha$ is an
  arbitrary vector field. } We assume there exists a background FLRW
metric $\bg_{\mu\nu}$ such that the difference $\g_{\mu\nu} -
\bg_{\mu\nu}$ can be seen as a ``small perturbation'' in the sense
discussed in Ref.~\cite{Vitenti2012}. Accordingly, we define the
tensor $\ddg_{\mu\nu}$ and its contravariant form as
\begin{align}\label{eq:def:ddg}
\ddg_{\mu\nu} \equiv \g_{\mu\nu} - \bg_{\mu\nu},\qquad
\ddg^{\mu\nu}\equiv\bg^{\mu \alpha}\bg^{\nu \beta}\ddg_{\alpha \beta}.
\end{align}
Here and in what follows, an overbar will always refer to a
background quantity.

We now consider the case of a general space-time foliation defined by
the normal timelike vector field $\n^\mu$ ($\n^\mu\n_\mu = -1$), whose
gradient can be decomposed into an extrinsic curvature
$\mathcal{K}_{\mu\nu}$ and an acceleration $a_\mu\equiv
\n^\sigma\nabla_\sigma\n_\mu$ through
\begin{equation}
\nabla_\mu \n_\nu = \mathcal{K}_{\mu\nu} -n_\mu a_\nu.
\label{Du}
\end{equation}
The projector orthogonal to $\n^\mu$ (induced metric on the spatial
hypersurfaces of the foliation) is given by
$$\h_{\mu\nu} = \g_{\mu\nu} + \n_\mu\n_\nu.$$  Its action on
an arbitrary tensor
$M_{\mu_1\dots\mu_m}^{\indspace\nu_1\dots\nu_k}$ is defined
as
$$\hp{M^{\indspace\nu_1\dots\nu_k}_{\mu_1\dots\mu_m}} \equiv
\h_{\mu_1}^{\phantom a \alpha_1}\dots\h_{\mu_m}^{\phantom
	a\alpha_m}\h_{\beta_1}^{\phantom a \nu_1}\dots\h_{\beta_k}^{\phantom
	a\nu_k}M^{\indspace\beta_1\dots\beta_k}_{\alpha_1\dots\alpha_m},
$$
thereby introducing the notation $\hp{\cdots}$, in terms of which the
extrinsic curvature is simply found to be $\mathcal{K}_{\mu\nu} \equiv
\hp{\nabla_\mu\n_\nu}$. The expansion factor $\EX$ and shear
$\sigma_{\mu\nu}$ are then given respectively by
\begin{equation}
\EX \equiv \mathcal{K}_\mu^{\ \mu} \ \ \ \hbox{and} \ \ \
\sigma_{\mu\nu} \equiv \mathcal{K}_{\mu\nu} -\frac13 \EX
\h_{\mu\nu}.
\end{equation}

An arbitrary tensor $N_{\mu_1\dots\mu_m}^{\indspace\nu_1\dots\nu_k}$
will be called spatial if it is invariant under the projection, i.e., if
it satisfies $\hp{N_{\mu_1\dots\mu_m}^{\indspace\nu_1\dots\nu_k}} =
N_{\mu_1\dots\mu_m}^{\indspace\nu_1\dots\nu_k}$. The covariant
derivative compatible with the spatial metric $\h_{\mu\nu}$ is
\begin{equation}
\scd_\alpha M_{\mu_1\dots\mu_m}^{\indspace\nu_1\dots\nu_k} \equiv
\hp{\cd_\alpha M_{\mu_1\dots\mu_m}^{\indspace\nu_1\dots\nu_k}}.
\end{equation}
This spatial covariant derivative defines the spatial Riemann curvature
tensor $\SR_{\mu\nu\alpha}{}^\beta$ as
\begin{equation}\label{eq:def:SR}
\left(\scd_\mu\scd_\nu - \scd_\nu\scd_\mu\right) A_\alpha \equiv
\SR_{\mu\nu\alpha}{}^\beta A_\beta,
\end{equation}
where $A_\beta=\gamma[A_\beta]$ is an arbitrary spatial vector field.
The spatial Laplace operator is represented by the symbol $\scd^2$,
i.e., $\scd^2 \equiv \scd_\mu\scd^\mu$. In what follows, we shall denote
the contraction with the normal vector field $\n^\mu$ with an index $\n$
(e.g., $M_{\alpha \n} \equiv M_{\alpha \beta}\n^\beta$).

The covariant derivative compatible with the background metric is
represented by the symbol $\bcd$ or by a semicolon, i.e.,
$\bg_{\mu\nu;\gamma} \equiv \bcd_\gamma\bg_{\mu\nu} = 0$. Using a
background foliation described by the normal vector field $\bn^\mu$, we
define the projector $\bh_{\mu\nu}$, spatial derivative $\bscd_\mu$ and
spatial Riemann tensor $\bSR_{\mu\nu\alpha}{}^\beta$, as we have done
for the objects derived from $\g_{\mu\nu}$. We use the symbol ``$
{}_{\scp} $'' to represent the background spatial derivative, i.e.,
$\bm{T}_{\scp\mu} \equiv \bscd_\mu \bm{T}$ for any tensor $\bm{T}$.

One should keep in mind that the background and perturbation tensors
have their indices lowered and raised always by the background metric.
Finally, we define the dot operator of an arbitrary tensor as
\begin{equation}
\dot{M}_{\mu_1\dots\mu_m}^{\indspace\nu_1\dots\nu_k} \equiv
\bhp{\lie_{\bn}{}M_{\mu_1\dots\mu_m}^{\indspace\nu_1\dots\nu_k}},
\end{equation}
where $\lie_{\bn}$ is the Lie derivative in the direction of
$\bn$.

Specifying to the case of an FLRW background, the extrinsic curvature
and the spatial Ricci tensor are diagonal and read
\[
\bEC_{\mu\nu} = \frac{1}{3}\bEX\bh_{\mu\nu},\qquad\bSR_{\mu\nu} =
2\bK\bh_{\mu\nu},
\]
with the expansion factor and the function $\bK$ being homogeneous,
i.e., $\bEX_{\scp\mu} = 0 = \bK_{\scp\mu}$. The expansion factor is
simply proportional to the Hubble function $H \equiv\frac13 \bEX$,
itself giving the rate of time evolution of the scale factor $a$, namely
$H\equiv \dot a/a$. Thus, the background Einstein tensor is given by
\begin{align}\label{eq:FLRW:ee}
\bG_{\mu\nu} &= \left(3\bK + \frac{1}{3}\bEX^2\right)\bn_\mu\bn_\nu -
\frac13 \left(3\bK+2\dot{\bEX} + \bEX^2\right)\bh_{\mu\nu}.
\end{align}

Having set the general notations, we now move to evaluate all the
relevant perturbative contributions, namely those due to geometry and
matter.

\subsubsection{The geometric contribution}

Let us define the ``true'' tensor \cite{Vitenti2013}
$\LL_{\mu\nu}{}^\alpha$ through the equation 
\begin{align}
&(\cd_\mu-\bcd_\mu)A_\nu = \LL_{\mu\nu}{}^\beta
  A_\beta,\nonumber\\ &\LL_{\alpha\beta}{}^\gamma =
  -\frac{1}{2}\g^{\gamma\sigma}\left(\g_{\sigma\beta;\alpha} +
  \g_{\sigma\alpha;\beta} -
  \g_{\alpha\beta;\sigma}\right)
 \label{eq:LL}
\end{align}
(since the same coordinate system $\{x^\mu\}$ is used for both the
background and the perturbed geometries, the partial derivatives cancel
out). With this tensor, we can now begin our evaluation of the geometric
contribution to the second order action by noting that since the
covariant derivative ``$\, ;\, $'' is with respect to the background
metric, we have, to first order,
\begin{equation}\label{eq:LL:1}
\LL_{\alpha\beta\gamma} \approx -\frac{1}{2}\left(\ddg_{\gamma\beta;\alpha}
+ \ddg_{\gamma\alpha;\beta} - \ddg_{\alpha\beta;\gamma}\right).
\end{equation}

The perturbed Riemann tensor is related to the background
Riemann tensor by the exact expression
\begin{equation}\label{eq:pR:R}
\R_{\mu\nu\alpha}{}^\beta = \bR_{\mu\nu\alpha}{}^\beta +
2\LL_{\alpha[\nu}{}^\beta{}_{;\mu]} +
2\LL_{\alpha[\mu}{}^\gamma\LL_{\nu]\gamma}{}^\beta,
\end{equation}
where $\bR_{\mu\nu\alpha}{}^\beta$ is the Riemann tensor
constructed with the background metric $\bg_{\mu\nu}$ and
the brackets represent antisymmetrization over the indices
bracketed.
The expansion of the curvature scalar up to second order is therefore
\begin{equation}\label{eq:pSR:SR}
\begin{split}
\R &\approx \bR + \bR_{\mu\nu}\xi^{\mu\nu} + \left(\xi^{\mu\nu}_{\ \ ;\nu}
- \xi^{;\mu}\right)_{;\mu} + \frac12 \xi_{;\mu} \left(\xi^{\mu\nu}_{\ \ ;\nu}
-\frac12\xi^{;\mu}\right) \\ &-
\LL_{\mu\nu\alpha}\LL^{\mu\alpha\nu} -
\xi^{\mu\nu}\left( \frac12\xi_{\nu;\mu}-\LL_{\mu\nu\ ;\sigma}^{\ \ \sigma} \right),
\end{split}
\end{equation}
where again $\bR$ and $\bR_{\mu\alpha}$ are, respectively,
the scalar curvature and the Ricci tensor of the background.
To complete the expansion of the Lagrangian, we also need the
metric determinant up to second order, namely
\begin{equation}
\sqrt{-\g} \approx \sqrt{-\bg}\left(1 + \frac12 \ddg -
  \frac14\ddg_{\mu\nu}\ddg^{\mu\nu} + \frac18\ddg^2\right).
\end{equation}

Given the background foliation $\bn^\mu$, the metric
perturbation can be decomposed as
\begin{equation}\label{eq:dgg:decomp}
\ddg_{\mu\nu} = 2\A\bn_\mu\bn_\nu+2\B_{(\mu}\bn_{\nu)}+2\C_{\mu\nu},
\end{equation}
where
\begin{align*}
\A \equiv \frac12 \ddg_{\bn\bn}, \quad \B_\mu \equiv -
\bhp{\ddg_{\bn\mu}},\quad \C_{\mu\nu} \equiv
\frac12\bhp{\ddg_{\mu\nu}}.
\end{align*}

Using the scalar, vector and tensor decomposition
(see~\cite{Stewart1990}), we rewrite the metric perturbations
as
\begin{align*}
\B_\mu &= \BS_{\scp\mu} + \BV_\mu, \\ \C_{\mu\nu} &= \CS\h_{\mu\nu} -
\CSD_{\scp\mu\nu} + \CV_{(\nu\scp\mu)} + \CTD_{\mu\nu},
\end{align*}
where $\BV^\mu{}_{\scp\mu} = \CV^\mu{}_{\scp\mu} =
\CTD_\mu{}^\nu{}_{\scp\nu} = \CTD_\mu{}^\mu = 0$. It is
straightforward to show (see the Appendix~A of
Ref.~\cite{Vitenti2013}) that, in terms of this decomposition,
the shear perturbation reads
\begin{equation}\label{eq:def:FLRW:sCI:decomp}
\dSH_{\mu\nu} =
\left[\bscd_{(\mu}\bscd_{\nu)}-\frac{\bh_{\mu\nu}\blap}{3}\right]\dSHs
+ \dSHv_{(\nu\scp\mu)} + \dot{\CTD}_\mu{}^\alpha\bh_{\alpha\nu},
\end{equation}
where we have defined
\begin{equation}\label{eq:FLRW:dSHs}
\dSHs \equiv \left(\BS-\dot{\CSD} + \frac{2}{3}\bEX\CSD\right)
\ \ \ \hbox{and} \ \ \
\dSHv^{\alpha} \equiv \BV^\alpha + \dot{\CV}^\alpha.
\end{equation}
The perturbation on the expansion factor gives
\begin{equation}\label{eq:FLRW:dEX}
\dEX = \blap\dSHs  + \bEX\A + 3\dot{\CS}.
\end{equation}
Finally, the perturbations on the spatial Ricci tensor and
curvature scalar are
\begin{align*}
\bhp{\dSR_\mu{}^\bn} =& 0, \\ \bhp{\dSR_\bn{}^\nu} =&
-2\bK \left(\BS^{\scp\nu} + \BV^\nu\right), \\ \bhp{\dSR_\mu{}^\nu} =&
-\CS_{\scp\mu}{}^{\scp\nu} - \bh_\mu{}^\n\left(\blap + 4\bK\right)\CS \\
 & -\left( \blap -
  2\bK\right)\CTD_\mu{}^\nu, \\
  \dSR =& -4\blapK\CS,
\end{align*}
where we have defined the operator $\blapK \equiv \blap + 3\bK$.

After all these definitions, the general expression for the
gravitational part of the second order Lagrangian is given
by $$\dLA^{(2)}_{\g} = \dLA^{(2,\text{s})}_\g + \dLA^{(2,\text{v})}_\g +
\dLA^{(2,\text{t})}_\g + \sqrt{-\bg}l^\tbg_\g,$$
where the scalar part reads
\begin{equation}\label{eq:FLRW:dLA2:s}
\begin{split}
&\frac{\dLA^{(2,\text{s})}}{\sqrt{-\bg}} = \frac{\blap\dSHs\blapK\dSHs}{3\kp}
  - \frac{\dEX^2}{3\kp}-2\left(\frac{\CS}{2}-\A\right) \frac{\blapK\CS}{\kp},
\end{split}
\end{equation}
while the vector and tensorial sectors are, respectively
\begin{align}\label{eq:FLRW:dLA2:v}
\frac{\dLA^{(2,\text{v})}_\g}{\sqrt{-\bg}} &=
\frac{\dSHv_{(\alpha\scp\nu)}\dSHv^{(\alpha\scp\nu)}}{2\kp},
\\ \label{eq:FLRW:dLA2:t} \frac{\dLA^{(2,\text{t})}_\g}{\sqrt{-\bg}} &=
\frac{\dot{\CTD}_\nu{}^\gamma\dot{\CTD}_\gamma{}^\nu +
  \CTD_\mu{}^\nu(\blap - 2K)\CTD_\nu{}^\mu}{2\kp},
\end{align}
as derived in Ref.~\cite{Vitenti2013}; note at this point that
the quantities $\dot\A$ and $\dot\BS$ do not appear in
the Lagrangian, having been transformed away through elimination
of a total derivative term.

Finally the terms proportional to the background Einstein
tensor read
\begin{equation}
\begin{split}
2\kp l^\tbg_\g = & \
\bG_{\bn\bn}\left(\B_\mu\B^\mu-\A^2-2\A\C\right)
\\ &+\bG_{\mu\nu}\bh^{\mu\nu}
\left(2\C^{\alpha\beta}\C_{\alpha\beta}
- \C^2\right).
\end{split}
\end{equation}

In what follows, after having written down the matter
contribution to the second order action, we shall
concentrate on the scalar parts of those, since it is the
only contribution which has, so far unambiguously, been
measured. Although the Lagrangian governing the scalar sector
of the perturbations depends on the four field variables
$(\A,\;\BS,\;\CS,\;\CSD)$,
it turns out to be more computationally efficient to
write it in terms of the kinematic quantities $\dEX$, $\dSH$
and $\dSR$, instead of directly in terms of the scalar
variables above.

\subsubsection{Matter part}

The second order Lagrangian for the matter part consisting
of a single generic fluid was also obtained in
Ref.~\cite{Vitenti2013}. We consider a fluid having
an energy density $\bed$, pressure $\bp$ and associated
sound velocity
$$\bcs^2 \equiv \frac{\del{}\bp}{\del\bed},$$
which is then perturbed so that the relevant
variables will be the energy density perturbation $\delta\rho$, the
pressure perturbation $\delta p$, and the velocity potential
$\vms$ of the normalized velocity field of the fluid
$\nm_\mu$ defined through
$$\nm_\mu = \bn_\mu - \A\bn_\mu + \vms_{\scp\mu}.$$

In general, there may also exist other
relevant quantities, like an intrinsic entropy perturbation
and other velocity potentials; here we will consider
multiple noninteracting barotropic fluids. In this case,
for each fluid, the intrinsic entropy perturbation and the
other degrees of freedom decouple from the energy and
pressure perturbations. We will therefore merely discard
them.

We now assume a system consisting of $N$ fluids,
labeled by latin indices.
As for the purely gravitational terms, the quantities $\dot\A$
and $\dot\BS$, which appear in the original form of the
Lagrangian, can be removed through elimination
of another total derivative term~\cite{Vitenti2013}. This
is possible for each fluid term independently.
The second order matter Lagrangian, according to
Eq.~(62) of Ref.~\cite{Vitenti2013}, thus reads
\begin{equation}\label{eq:def:dLA2:m}
\begin{split}
\frac{\dLA_{\mat{}i}^{(2)}}{\sqrt{-\bg}} =&
  \frac{\bcsi{i}^2(\dedgi_{i})^2}{2\bedpi{i}} +
  \frac12\bedpi{i}\vmsi{i}\blapK\vmsi{i} \\ & -
  \frac{3\kp}{4}\bedpi{i}^2\vmsi{i}^2 + \bedpi{i}\dEX\vmsi{i} +
  l^\tbg_{\mat{}i} \\ &- \frac{\bedi{i}}{2}\left(\B_\gamma\B^\gamma -
  \A^2 - 2\C\A\right) \\ &-
  \frac{\bpi{i}}{2}\left(2\C_\gamma{}^\nu\C_\nu{}^\gamma -
  \C^2\right),
\end{split}
\end{equation}
where the term $l^\tbg_{\mat{}i}$ is given by
\begin{equation}
\begin{split}
l^\tbg_{\mat{}i} =& -\frac34\bedpi{i}\vmsi{i}^2\left(\bG_{\bn\bn}
- \kp\bedi{i}\right)
\\ &-\frac14\bedpi{i}\vmsi{i}^2\left(\bG_{\mu\nu}\bh^{\mu\nu}
- 3\kp\bpi{i}\right),
\end{split}
\end{equation}
and
\begin{align}
\dedgi_{i} &= \ded_{i} - \bEX\bedpi{i}\vmsi{i}, \\
\ded_{i} &= -\bedpi{i}\frac{\vmsidt{i} -
\bcsi{i}^2\bEX\vmsi{i} - \A}{\bcsi{i}^2}.
\label{drhoi}
\end{align}
It is worth noting that in the case of a barotropic fluid
set, each fluid introduces only one scalar field variable
to the problem, namely $\vmsi{i}$, so that the whole
matter Lagrangian depends on the set of variables $\{\vmsi{i}\}_N$.
As in the gravitational sector case, it is more convenient
to write the Lagrangian in terms of the relevant
physical quantities. For each fluid, this is the
energy density perturbation $\ded_{i}$ and its gauge-invariant
version $\dedgi_{i}$. This is the only possible (apart from an
irrelevant rescaling) gauge invariant combination that one
can form using variables of the same fluid, namely $\vmsi{i}$
and $\vmsidt{i}$.

Here we have discarded all terms proportional to the background
equations for each fluid separately. This is possible because the fluids
are not coupled, which means that the equation of motion for each fluid
appears in the first order Lagrangian. Therefore, as explained in
Ref.~\cite{Vitenti2013}, we can remove such terms from the Lagrangian
with a mere redefinition of the perturbation variables. This is done by
modifying the second order by summing products of the first order
variables. Since the first order Lagrangian appears as the product of
the background equations of motion times the perturbations, this
redefinition leaves the first order Lagrangian unmodified but introduces
new second order terms in the Lagrangian, always multiplied by the
zeroth order equations of motions, which are used to cancel out these
unwanted terms. In the case at hand however, we are forced to keep the
terms $l^\tbg_{\mat{}i}$ because the Einstein equations appearing in the
first order Lagrangian involve the total energy density and pressure,
and not individual ones. For this reason, we rewrite each
$l^\tbg_{\mat{}i}$ as
\begin{equation}
\begin{split}
l^\tbg_{\mat{}i} =& \frac{3\kappa}{4}\bedpi{i}^2\vmsi{i}^2 -
\frac{3\kappa}{4}\bedpi{i}(\bed+\bp)\vmsi{i}^2
\\ &-\frac34 \bedpi{i} \vmsi{i}^2\left(\bG_{\bn\bn} - \kp\bed\right)
\\ &-\frac14 \bedpi{i} \vmsi{i}^2\left(\bG_{\mu\nu}\bh^{\mu\nu}
- 3\kp\bp\right),
\end{split}
\label{eq:lbgmi}
\end{equation}
where the total energy density and pressure are,
\begin{equation}
\bed = \sum_{i}\bedi{i},\qquad\bp = \sum_{i}\bpi{i},
\label{bed}
\end{equation}
and we note $\kappa=8\pi\GN = \MP^{-2}$ (in units where
$\hbar = c = 1$). In Eq.~\eqref{bed} and the following, we assume
all sums, unless explicitly stated otherwise, to run from 1 to $N$, 
the total number of fluids considered.

Now, we can remove the terms in the last two lines of 
Eq.~\eqref{eq:lbgmi} using the first order gravitational Lagrangian, 
and add the first line explicitly to the matter Lagrangian, yielding
\begin{equation}\label{eq:dLA2:m}
\begin{split}
\frac{\dLA_{\mat{}i}^{(2)}}{\sqrt{-\bg}} =&
  \frac{\bcsi{i}^2(\dedgi_{i})^2}{2\bedpi{i}} +
  \frac{\bedpi{i}}{2}\vmsi{i}\blapK\vmsi{i} \\ &-
  \frac{3\kp}{4}\bedpi{i}(\bed + \bp)\vmsi{i}^2 +
  \bedpi{i}\dEX\vmsi{i} \\ &-
  \frac{\bedi{i}}{2}\left(\B_\gamma\B^\gamma - \A^2 - 2\C\A\right) \\ &-
  \frac{\bpi{i}}{2}\left(2\C_\gamma{}^\nu\C_\nu{}^\gamma -
  \C^2\right),
\end{split}
\end{equation}
so that the total Lagrangian will be given by the sum
$$
\frac{\dLA^{(2)}}{\sqrt{-\bg}} =
\frac{\dLA_{\g}^{(2)}}{\sqrt{-\bg}} +
\sum_i\frac{\dLA_{\mat{}i}^{(2)}}{\sqrt{-\bg}}.
$$
Using Eqs.~\eqref{eq:FLRW:dLA2:s} and \eqref{eq:dLA2:m}, we obtain
\begin{equation}\label{eq:FLRW:dLA2}
\begin{split}
\frac{\dLA^{(2,\text{s})}}{\sqrt{-\bg}} =& \frac{\blap\dSHs\blapK\dSHs}{3\kp}
  - \frac{\dEX^2}{3\kp} - 2\left(\CS-\A\right) \frac{\blapK\CS}{\kp}
  \\ &+\sum_{i}\left[\frac{\bcsi{i}^2(\dedgi_{i})^2}{2\bedpi{i}}
    + \frac12\bedpi{i}\vmsi{i}\blapK\vmsi{i}\right]
  \\ &+\sum_{i}\left[\bedpi{i}\dEX\vmsi{i}-\frac{3\kp}{4}\bedpi{i}(\bed
    + \bp)\vmsi{i}^2\right],
\end{split}
\end{equation}
where we have again removed the terms linear in the
background field equations using the first order Lagrangian.

Multiplying and dividing the term containing $\dEX^2$ by
$(\bed+\bp)$, and rewriting the numerator as a sum of
$\bedpi{i}$, we can combine it with the last line of the
above, obtaining
\begin{equation}\label{eq:FLRW:dLA2:tmp}
\begin{split}
\frac{\dLA^{(2,\text{s})}}{\sqrt{-\bg}} =& \frac{\blap\dSHs\blapK\dSHs}{3\kp}
  +\left(2\A-\CS\right) \frac{\blapK\CS}{\kp}
  \\ &+\sum_{i}\left[\frac{\bcsi{i}^2(\dedgi_{i})^2}{2\bedpi{i}}
    + \frac12\bedpi{i}\vmsi{i}\blapK\vmsi{i}\right]
  \\ &-\frac{1}{3\kp(\bed+\bp)}\sum_{i}\bedpi{i}\left[\dEX-
  \frac{3\kp}{2}(\bed+\bp)\vmsi{i}\right]^2.
\end{split}
\end{equation}

It is convenient to work with gauge-invariant variables as
e.g., $\dedgi_{i}$, in terms of which we can express all other
quantities. First, we define a gauge-invariant expansion factor
\begin{equation}
\dEXi \equiv \dEX - \frac{3\kp}{2}(\bed+\bp)\vms -
3 \frac{\blapK\CS}{\bEX},
\end{equation}
where we introduced the total velocity potential $\vms$
\begin{equation}
\vms \equiv \frac{1}{\bed + \bp}\sum_{i}\bedpi{i}\vmsi{i}.
\end{equation}
This potential represents the weighted average of the velocity
potentials for each fluid using $\bedpi{i}$ as its weight.

Substituting this expression back into the total Lagrangian yields
\begin{equation}\label{eq:FLRW:dLA2:cmp}
\begin{split}
&\frac{\dLA^{(2,\text{s})}}{\sqrt{-\bg}} =
  \frac{3\blap\CSI\blapK\CSI}{\kp\bEX^2} - \frac{\dEXi^2}{3\kp} + \sum_{i=1}^n\frac{\bcsi{i}^2(\dedgi_{i})^2}{2\bedpi{i}}\\
  &-\sum_{i}\frac{9\bedpi{i}}{2\bEX^2}\left[\frac{3\kp\bedp}{2}(\CSvms-\CSvmsi{i})^2 - \CSvmsi{i}\blapK\CSvmsi{i}\right],
\end{split}
\end{equation}
where we have used the gauge-invariant variables
\begin{equation}
\CSvmsi{i}
\equiv \CS + \frac13\bEX\vmsi{i} \qquad \hbox{and} \qquad
\CSI \equiv \CS -
\frac13\bEX\dSHs.
\label{eq:def:CSvmsi}
\end{equation}
The variable $\CSvms$ is the weighted average
of $\CSvmsi{i}$, analogous to $\vms$. We have also removed
a surface term and a linear term proportional to the
background equations of motion, following what was done
in Eq.~(69) of Ref.~\cite{Vitenti2013}. Note that, in these
variables, the Lagrangian in Eq.~\eqref{eq:FLRW:dLA2:cmp}
naturally reduces to Eq.~(69) of~\cite{Vitenti2013} when
only a single fluid is present.

The final Lagrangian in Eq.~\eqref{eq:FLRW:dLA2:cmp} is
organized as usual, the first $2+N$ kinematic terms involving
the squares of the time derivatives and the last $N$ terms 
the fluid potentials. It is a functional of the $4+N$ field 
variables $(\A,\;\BS,\;\CS,\;\CSD,\;\{\vmsi{i}\}_N)$.
As was done for the individual matter and gravitational parts,
here we also wrote the Lagrangian in term of physical quantities
instead of directly in terms of these variables. This choice
of variables simplifies the manipulation of the different terms
in the Lagrangian (compare, for example, with the manipulations done
in~\cite{Mukhanov1992}). Nonetheless, the most important aspect
of this form is the way it simplifies the constraints reduction
as we see in the next section.

\subsection{Hamiltonian}
\label{sec:H}

The Hamiltonian formulation depends on the Legendre transform
of the Lagrangian with respect to the field time derivatives.
This transform is possible when the Hessian matrix is
nonsingular, which is not the case here. For a singular
Hessian one can use the
Faddeev--Jackiw~\cite{Faddeev1988,Jackiw1993}
procedure. It starts by identifying the null-space of the
Hessian matrix. The Lagrangian~\eqref{eq:FLRW:dLA2:cmp}
depends on $4+N$ field variables and only on the time
derivatives of $2+N$ variables, namely
$(\CS,\;\CSD,\;\{\vmsi{i}\}_N)$.
This allows an automatic identification of $\A$ and $\BS$ as
part of the null-space of the Hessian matrix\footnote{It is
worth noting that the identification of the null-space is not
straightforward as in a generic linear algebra problem. Any
time dependent linear point transformation of the field
variables generates additional terms depending on the time
derivative of the transformation matrix.}. The time derivatives
of $\CS$ and $\CSD$ are contained in $\CSI$ and $\dEXi$
respectively through $\dSHs$ and $\dEX$. However, $\dEX$
depends on both $\dot{\CSD}$ and $\dot{\CS}$ while $\dSHs$
depends only on $\dot{\CSD}$ [see Eqs.~\eqref{eq:FLRW:dSHs}
and \eqref{eq:FLRW:dEX}]. We can disentangle the two variables
using,instead of $\CS$, the field
\begin{equation}
\CST \equiv \frac{\C}{3} = \CS - \frac{\blap\CSD}{3},
\end{equation}
where $\C$ is the trace of the spatial projection of the
metric perturbation in Eq.~\eqref{eq:dgg:decomp}.
In terms of this variable, the perturbation
on the expansion rate then takes the simpler form
$$\dEX = 3\dot{\CST} + \blap\BS + \A\bEX.$$

Given the above, we are in a position to calculate the canonically
conjugate momenta to $\dot{\CST}, \lie_\bn{\bscd^2\CSD}$ and
$\vmsidt{i}$. They read
\begin{align}\label{eq:def:pCSD}
\pvmsi{i} &\equiv \frac{\del\dLA^{(2,\text{s})}}{\del\vmsidt{i}} = -\sqrt{-\bg}\dedgi_{i}, \\ \label{PiEps}
\pDCSD &\equiv\frac{\del\dLA^{(2,\text{s})}}{\del{\lie_\bn(\blap\CSD)}} =
\frac{2\sqrt{-\bg}}{\kp\bEX}\blapK\CSI,
\\ \label{PiPsit} \pCST &\equiv \frac{\del\dLA^{(2,\text{s})}}{\del\dot{\CST}} = -
2\sqrt{-\bg}\frac{\dEXi}{\kp}.
\end{align}
Since we arranged the Lagrangian~\eqref{eq:FLRW:dLA2:cmp} such
that each kinematic term depends only on the time
derivative of a single variable, each equation above relates
one momentum with only one time derivative.

Solving the expressions (\ref{eq:def:pCSD}--\ref{PiPsit}) in terms of the time derivatives,
we obtain
\begin{align} \label{eq:dotvmsi}
&\vmsidt{i} = \frac{\bcsi{i}^2\pvmsi{i}}{\sqrt{-\bg}\bedpi{i}} + \A, \\ \label{eq:dotU}
&\lie_\bn\left(\blap\CSD\right) =  \frac{3\kp\blap\bilapK\pDCSD}{2\sqrt{-\bg}}
 - \frac{3\blap\CS}{\bEX}+ \blap\BS, \\ \label{eq:dotCST}
&\dot{\CST} = - \frac{\kp\pCST}{6\sqrt{-\bg}}
  +\frac12\kp(\bed + \bp)\vms
  + \frac{\blapK\CS}{\bEX} - \frac{\bEX}{3}\A - \frac{\blap\BS}{3},
\end{align}
and performing a Legendre transform in these variables then yields
\begin{equation}\label{eq:La:const}
\dLA^{(2,\text{s})} = \pDCSD\lie_\bn({\blap\CSD}) + \pCST\dot{\CST} +
\sum_{i}\pvmsi{i}\vmsidt{i}- \dHA_\text{c}^{(2,\text{s})},
\end{equation}
where
\begin{equation}\label{eq:HA:1}
\begin{split}
&\dHA_\text{c}^{(2,\text{s})} = \pDCSD\lie_\bn({\blap\CSD}) + \pCST\dot{\CST} +
  \sum_{i}\pvmsi{i}\vmsidt{i} \\
  &-\frac{3\kp\pDCSD\blap\bilapK\pDCSD}{4\sqrt{-\bg}} +
  \frac{\kp\pCST^2}{12\sqrt{-\bg}} -
  \sum_{i}\frac{\bcsi{i}^2\pvmsi{i}^2}{2\sqrt{-\bg}\bedpi{i}}\\
  &+\sum_{i}\frac{9\sqrt{-\bg}\bedpi{i}}{2\bEX^2}
    \left[\frac{3\kp\bedp}{2}(\CSvms-\CSvmsi{i})^2
      - \CSvmsi{i}\blapK\CSvmsi{i}\right],
\end{split}
\end{equation}
is the constrained Hamiltonian we are looking for, assuming time
derivatives appearing in the first line to be given by
Eqs.~\eqref{eq:dotvmsi}~--~\eqref{eq:dotCST}. Note that the Hamiltonian
is a functional of the $2N+4$ dimensional phase space fields
$(\{\vmsi{i},\;\pvmsi{i}\}_N,\;\CST,\;\pCST,\;\CSD,\;\pDCSD)$, and also
depends on $\A$ and $\BS$. However, this dependence is linear and only
through the first three terms in the Hamiltonian~\eqref{eq:HA:1} [see
Eqs.~\eqref{eq:def:pCSD}--\eqref{PiPsit}]. In other words, $\A$ and
$\BS$ act as Lagrange multipliers so that variations of the Hamiltonian
with respect to these variables provide the relevant constraints on the
phase space.

Varying the Hamiltonian with respect to $\A$ and $\BS$ respectively
yields
\begin{equation}\label{eq:const:1}
\frac{\del\dHA_\text{c}^{(2,\text{s})}}{\del\A} = \sum_{i}\pvmsi{i}
- \frac{\bEX}{3}\pCST = 0,
\end{equation}
and
\begin{equation}\label{eq:const:2}
\frac{\del\dHA_\text{c}^{(2,\text{s})}}{\del\blap\BS} = \pDCSD - \frac{\pCST}{3} = 0.
\end{equation}
These two equations reduce the number of momenta by two. We choose
to to express $\pCST$ and $\pDCSD$ in terms of $\{\pvmsi{i}\}_N$, which,
applied to the kinematic part of the Lagrangian~\eqref{eq:La:const}
represented by the first $2+N$ terms, yields
$$
\dLA^{(2,\text{s})} = \sum_{i}\pvmsi{i}\left(\frac{3}{\bEX}\dot{\CS} +
\vmsidt{i}\right) - \dHA_\text{c}^{(2,\text{s})},
$$
and we are left with only $N$ momenta. Given the 
Lagrangian structure, when we reduce the momenta through linear 
constraints, the field variables containing time derivatives will
also appear as some linear combination induced by the new momenta.
In the equation above, the momenta multiply the gauge invariant 
variable $ \CSvmsi{i} $ [Eq.~\eqref{eq:def:CSvmsi}]. Rewriting 
the terms in the parenthesis above generates new terms,
which must be included in the final Hamiltonian, i.e.,
$$
\dLA^{(2,\text{s})} = \sum_{i}\pCSvmsi{i}\CSvmsidt{i} -
\sum_{i}\frac{\dot{\bEX}}{\bEX}\pCSvmsi{i}\CSvmsi{i} +
\frac{\dot{\bEX}}{\bEX}\pCSvms\CS - \dHA_\text{c}^{(2,\text{s})},
$$
where we defined new momenta
$$
\pCSvmsi{i} \equiv \frac{3}{\bEX}\pvmsi{i},
$$
and the total momentum
$$
\pCSvms \equiv \sum_{i}\pCSvmsi{i}.
$$
Finally, we can group the terms in the Lagrangian above and define the
unconstrained Hamiltonian as
\begin{equation}
\dHA^{(2,\text{s})} \equiv \left.\dHA_c^{(2,\text{s})}\right\vert_{\substack{\pCST = \pCSvms\\\pDCSD = \frac13\pCSvms}}
  + \sum_{i}\frac{\dot{\bEX}}{\bEX}\pCSvmsi{i}\CSvmsi{i}
  - \frac{\dot{\bEX}}{\bEX}\pCSvms\CS.
\end{equation}
Arranging these terms we write the final Lagrangian and Hamiltonian
explicitly
\begin{equation}\label{eq:LA:final}
\dLA^{(2,\text{s})} = \sum_{i}\pCSvmsi{i}\CSvmsidt{i} - \dHA^{(2,\text{s})},
\end{equation}
where the Hamiltonian, after removing another term proportional to a
background equation, reads
\begin{equation}\label{eq:HA:final}
\begin{split}
&\dHA^{(2,\text{s})} =
  \sum_{i}\frac{\dot{\bEX}}{\bEX}\pCSvmsi{i}\CSvmsi{i} +
  \frac{3\kp(\bed + \bp)}{2\bEX}\CSvms\pCSvms \\
  &+\sum_{i}
    \frac{\bcsi{i}^2\bEX^2\pCSvmsi{i}^2}{18\sqrt{-\bg}\bedpi{i}}
    -\frac{\kp\bK\pCSvms\bilapK\pCSvms}{4\sqrt{-\bg}}\\
  &+\sum_{i}\frac{9\sqrt{-\bg}\bedpi{i}}{2\bEX^2}
    \left[\frac{3\kp\bedp}{2}(\CSvms-\CSvmsi{i})^2
      - \CSvmsi{i}\blapK\CSvmsi{i}\right].
\end{split}
\end{equation}

The Lagrangian~\eqref{eq:LA:final} shows that there are $2N$ dynamical
variables represented by the gauge invariant velocity potential and
their momenta, i.e., $\{\CSvmsi{i}, \pCSvmsi{i}\}_N$. Naturally, the
two scalar variables, $\A$ and $\B$, are absent from the unconstrained
Hamiltonian~\eqref{eq:HA:final}. Since there are no momenta associated
to $\CSD$ and $\CS$, their appearance in the Hamiltonian would lead
to new constraints, and indeed, they automatically cancel out.

This choice of variables satisfies the constraints and reduces the total
number of variables to $2N$. Using the two constraints
\eqref{eq:const:1} and \eqref{eq:const:2}, we can recover the two
momenta $\pCSD$ and $\pCST$. Equations \eqref{eq:dotU} and
\eqref{eq:dotCST}, relating the time derivatives
and the momenta, can then be
integrated to provide the values for the missing variables
$(\A,\;\BS,\;\CSD,\;\CS).$ Naturally, one must also add two gauge
conditions to completely fix this set of equations.

The form \eqref{eq:HA:final} decouples at leading order in the short
wavelength limit. Despite this fact, it is however not as convenient, as
will be shown explicitly in Sec.~\ref{sec:Q}, for the subsequent
quantization, because particle creation, influenced by the
next-to-leading order terms, leads to nonconvergent $\beta$ functions.
One could change variables directly to a more convenient set, but we
present in the following section an intermediate step, thus introducing
the equivalent, in the single-fluid case, of the Mukhanov-Sasaki
variable. This step permits to write the Hamiltonian in a block-diagonal
form.

\subsection{Adiabatic and Entropy Splitting}
\label{sec:split}

Our way of solving the constraints introduced new terms in the
Hamiltonian~\eqref{eq:HA:final}, involving the total gauge invariant
velocity potential and its momentum. These terms couple all
dynamical variables. Varying the Hamiltonian with respect
to $\CSvmsi{i}$ or $\pCSvmsi{i}$ yields terms proportional to $\CSvms$
and/or $\pCSvms$ which are related to the adiabatic modes of the
perturbations. Thus, we can make a series of canonical transformations
to rewrite the system in terms of adiabatic and entropy modes. In what
follows, we perform these transformations, exhibit the adiabatic and
entropy expansion for the general $N$-fluid case, and apply our result
to the simpler (but natural) situation in which only two fluids, e.g.
radiation and dust, are present.

\subsubsection{General case: $N$ fluids}

First, we rewrite $\{\CSvmsi{i},\;\pCSvmsi{i}\}_N$ in
terms of $(\CSvms,\;\pCSvms)$ and the relative variables
\begin{equation}
\CSvmsmi{i} \equiv \CSvmsi{i} - \CSvms\qquad \hbox{and}\qquad\pCSvmsmi{i} \equiv
\pCSvmsi{i} - \frac{\bedpi{i}}{(\bed+\bp)}\pCSvms.
\end{equation}
These variables are not independent since
$\sum_{i}\pCSvmsmi{i} = 0$ and $\sum_{i}\bedpi{i}\CSvmsmi{i} = 0$.
In terms of those, the total Lagrangian reads
\begin{equation}
\begin{split}
\dLA^{(2,\text{s})} =& \sum_{i}\pCSvmsmi{i}\CSvmsmidt{i} + \pCSvms\dot{\CSvms} \\
&- \dHA^{(2,\text{s})} +
\frac{\bEX\pCSvms}{\bed+\bp}\sum_{i}\bedpi{i}\bcsi{i}^2\CSvmsmi{i}.
\end{split}
\end{equation}

Defining the total (weighted average) sound speed
\begin{equation}\label{eq:def:bcs}
\bcs^2 \equiv \frac{1}{\bed + \bp}\sum_{i}\bedpi{i}\bcsi{i}^2,
\end{equation}
and the new gauge-invariant variable $\MS$ (the usual curvature perturbation)
\begin{equation}
\MS \equiv \CSvms -\frac{\bK\bEX\bilapK\pCSvms}{3\sqrt{-\bg}\bedp},
\end{equation}
closely related to the Mukhanov-Sasaki variable, we can define a new
Hamiltonian as
$$
\dHA^{(2,\text{s})}_\mathrm{n} \equiv \dHA^{(2,\text{s})} -
  \frac{\bEX\pCSvms}{\bed+\bp}\sum_{i}\bedpi{i}\bcsi{i}^2\CSvmsmi{i},
$$
which is, explicitly
\begin{equation}
\begin{split}
&\dHA^{(2,\text{s})}_\mathrm{n} = - \frac{\kp\bK\pCSvms\bilapK\pCSvms}{4\sqrt{-\bg}}
  -\frac{\bK^2\pCSvms\bilapK\pCSvms}{2\sqrt{-\bg}\bedp} \\
  &+\frac{\bcs^2\bEX^2\pCSvms^2}{18\sqrt{-\bg}\bedp} -
  \sqrt{-\bg}\frac{9\bedp}{2\bEX^2}\MS\blapK\MS \\
  &+\sum_{i}\Bigg\{\frac{\dot{\bEX}}{\bEX}\pCSvmsmi{i}\CSvmsmi{i} +
    \frac{27\sqrt{-\bg}\kp(\bed+\bp)}{4\bEX^2}\bedpi{i}\CSvmsmi{i}^2 \\
  &+\frac{\bcsi{i}^2\bEX^2\pCSvmsmi{i}^2}{18\sqrt{-\bg}\bedpi{i}}-
    \sqrt{-\bg} \frac{9\bedpi{i}}{2\bEX^2}\CSvmsmi{i}\blapK\CSvmsmi{i}\\
  &-\frac{\bEX\pCSvms}{3\bedp}\bcsi{i}^2\bedpi{i}\left[3\CSvmsmi{i}-
    \frac{\bEX\pCSvmsmi{i}}{3\sqrt{-\bg}\bedpi{i}}\right]\Bigg\},
\end{split}
\end{equation}
in which yet another term proportional to the background equations
of motion was removed.  Finally, one can identify the term
$$\pCSvms \lie_\bn \left[
\frac{\bK\bEX\bilapK\pCSvms}{3\sqrt{-\bg} \bedp} \right] \subset
\dHA^{(2,\text{s})}_\mathrm{n}$$
in the above, which can be combined with other terms to yield a total
derivative that can subsequently safely be taken out. As a result,
one finally arrives at the matter Lagrangian in the form of
\begin{equation}
\begin{split}
\dLA^{(2,\text{s})} &= \dLA^{(2,\text{s})}_\text{a} + \dLA^{(2,\text{s})}_\en,
\end{split}
\end{equation}
which represents the split, as expected: the first part,
$\dLA^{(2,\text{s})}_\text{a}$ is the usual Lagrangian for the
adiabatic degree of freedom,
\begin{align}
\dLA^{(2,\text{s})}_\text{a} &= \pMS\dot{\MS} - \dHA_\text{a}^{(2,\text{s})},
\\ \label{eq:HA:adia} \dHA^{(2,\text{s})}_\text{a} &=
\frac{\bcs^2\bEX^2\pMS\blap\bilapK\pMS}{18\sqrt{-\bg}\bedp} -
\sqrt{-\bg}\frac{9\bedp}{2\bEX^2}\MS\blapK\MS,
\end{align}
where the momentum conjugate
to $\MS$ is $\pMS \equiv \pCSvms$, while
the entropy degrees of freedom are governed by the
second part
\begin{equation}
\label{lagrangian-s}
\dLA^{(2,\text{s})}_\en = \sum_{i}\pCSvmsmi{i}\CSvmsmidt{i} -
\dHA_\en^{(2,\text{s})},
\end{equation}
with the entropy Hamiltonian being
\begin{equation}
\begin{split}
\dHA^{(2,\text{s})}_\en =&
  \sum_{i}\Bigg[\frac{\dot{\bEX}}{\bEX}\pCSvmsmi{i}\CSvmsmi{i}
    + \frac{27\sqrt{-\bg}\kp(\bed+\bp)}{4\bEX^2}\bedpi{i}\CSvmsmi{i}^2
    \\ &+
    \frac{\bcsi{i}^2\bEX^2\pCSvmsmi{i}^2}{18\sqrt{-\bg}\bedpi{i}}
    -\sqrt{-\bg}
    \frac{9\bedpi{i}}{2\bEX^2}\CSvmsmi{i}\blapK\CSvmsmi{i} \\ &-
    \frac{\bEX\pMS}{3\bedp}\bcsi{i}^2\bedpi{i}\dedcmi{i}\Bigg],
\end{split}
\label{hamiltonian-s}
\end{equation}
where we have defined the gauge-invariant energy density contrast
\begin{equation}
\dedci{i} =
3\CSvmsi{i}-\frac{\bEX\pCSvmsi{i}}{3\sqrt{-\bg}\bedpi{i}} =
\frac{\ded_{i}}{\bedi{i}+\bpi{i}} + 3\CS,
\label{dpsi}
\end{equation}
and $\ded_{i}$ is given by Eq.~\eqref{drhoi}.
We can also define the variables
$$ \dedc =
\frac{1}{\bedp}\sum_{i}\bedpi{i}\dedci{i}\qquad\hbox{and}\qquad
\dedcmi{i} = \dedci{i} - \dedc $$ for later convenience.  Note that the
only coupling with the adiabatic degree of freedom appears in the last
line of the Hamiltonian
\eqref{hamiltonian-s}.

Expressing the Lagrangian \eqref{lagrangian-s} as a function of
$\dedcmi{i}$, one gets
\begin{equation}
\dLA^{(2,\text{s})}_\en =
\sum_i\frac{3\sqrt{-\bg}\bedpi{i}\CSvmsmi{i}}{\bEX}{\dedcmidt{i}} -
\dHA_\en^{(2,\text{s})},
\label{Lag2}
\end{equation}
where now
\begin{equation}
\begin{split}
\dHA^{(2,\text{s})}_\en =&
  \sum_{i}\Bigg[\frac{\sqrt{-\bg}\bcsi{i}^2\bedpi{i}\dedcmi{i}^2}{2}
    \\ &-
    \sqrt{-\bg}\frac{9\bedpi{i}}{2\bEX^2}\CSvmsmi{i}\blap\CSvmsmi{i} \\ &-
    \frac{\bEX\pMS}{3\bedp}\bcsi{i}^2\bedpi{i}\dedcmi{i}\Bigg].
\end{split}
\label{lagrangian-s2}
\end{equation}

In the final Hamiltonian \eqref{lagrangian-s2}, we moved from the
original variables $\{ \CSvmsi{i},\pCSvmsi{i}\}$ to the adiabatic split
involving the curvature perturbation $\{ \MS,\pMS\}$ and the entropy
modes $\{ \CSvmsmi{i},\pCSvmsmi{i}\}$; at this stage, it seems we have
$N+1$ mode variables and momenta ($N$ entropy and one adiabatic),
although we started with $N$. This comes from the fact that we still
have the constraints
\begin{equation}
\sum_{i}\bedpi{i}\CSvmsmi{i} = 0 = \sum_{i}\bedpi{i}\dedcmi{i}
\label{vinculos}
\end{equation}
to implement. Explicitly, we defined a set of new variables $\tilde P_i
\equiv -\dedcmi{i}$ and $\tilde Q_i \equiv 3\sqrt{-\bg} \bEX^{-1}
\bedpi{i} \CSvmsmi{i}$, in terms of which the Lagrangian \eqref{Lag2}
reads
\begin{equation}
\dLA^{(2,\text{s})}_\en = - \sum_i \tilde Q_i \dot{\tilde P}_i -
\dHA_\en^{(2,\text{s})} \ \ \ \to \ \ \ \sum_i \tilde P_i \dot{\tilde Q}_i -
\dHA_\en^{(2,\text{s})},
\label{Lag3}
\end{equation}
where we have discarded the total derivative term. At this stage, one
needs to set a reference fluid, indexed $\ell$ say, with respect to which
the relevant degrees of freedom will be defined. We thus define
$\tilde P_{i,\ell} \equiv \tilde P_i -\tilde P_\ell$, leading to
\begin{equation}
\dLA^{(2,\text{s})}_\en = \sum_{i\not= \ell} \tilde P_{i,\ell} \dot{\tilde Q}_i -
\dHA_\en^{(2,\text{s})},
\label{Lag4}
\end{equation}
showing that, indeed, the actual number of degrees of freedom is $N-1$.

Equations~\eqref{vinculos} permit us to rewrite the dependent variables $\tilde P_\ell$ and
$\tilde Q_\ell$ in terms of the $N-1$ independent ones explicitly through
\begin{equation}
\tilde P_\ell = - \sum_{i\not= \ell} \frac{\bedpi{i}}{\bedp} \tilde P_{i,\ell} \ \ \ 
\hbox{and} \ \ \ \tilde Q_\ell = - \sum_{i\not= \ell} \tilde Q_i.
\label{vinc2}
\end{equation}
The resulting total  Hamiltonian, whose properties we discuss thoroughly
in Sec.~\ref{sec:vevol}, includes \eqref{eq:HA:adia} and
\begin{equation}
\begin{split}
\dHA_\en^{(2,\text{s})} = \sum_i \bigg[ & \frac12\sqrt{-\bg} \bcsi{i}^2
\bedpi{i} \tilde P_i^2  - \frac{\tilde Q_i\blap \tilde Q_i}{2\sqrt{-\bg} \bedpi{i}} \\
 &+ \frac{\bEX \pMS}{3\bedp} \bcsi{i}^2 \bedpi{i}\tilde P_i \bigg];
\end{split}
\label{HsTOT}
\end{equation}
this Hamiltonian couples all the relevant field variables and all the corresponding
momenta, without mixing the variables and the momenta, i.e. it does not include
terms of the form $\propto \tilde Q_i \tilde P_{j,\ell}$, regardless of $i$ and $j$,
in contrast with our original Hamiltonian \eqref{eq:HA:final}.

\subsubsection{The 2-fluid case}

We now consider the step discussed above for the special case of a
two-fluid model for which no privileged fluid need be defined. In this
case, a simple expansion in terms of one adiabatic ($\{ \MS, \pMS\}$)
and one entropy ($\{Q, P_Q\}$) mode
is well-defined, which permits an easy writing of the Hamiltonian as
well as, as it will turn out in Sec.~\ref{sec:condini}, a natural way to
set vacuum initial conditions. 

Applying straightforwardly the scheme developed above, one sets $\ell=2$, leading to
the definition of the single momentum degree of freedom $P_Q$ through
\begin{equation}
\begin{split}
P_Q \equiv \tilde P_{1,2} = \tilde P_1 - \tilde P_2 &=
\dedcmi{2} - \dedcmi{1} \\
&= \frac{\ded_2}{\bedpi{2}} -\frac{\ded_1}{\bedpi{1}}
\end{split}
\end{equation}
[see Eq.~\eqref{dpsi}], and the conjugate variable
\begin{equation}
Q\equiv \tilde Q_1 =\frac{3\sqrt{-\bg}}{\bEX} \bedpf \left( \CSvmsi{1} - \CSvmsi{2}\right)
\end{equation}
where
\begin{equation}
\bedpf \equiv \frac{\bedpi{1}\bedpi{2}}{\bedp}.
\end{equation}
The other relevant variables can all be expressed in terms of those above, namely
\begin{equation}
\tilde P_1 = \frac{\bedpi{2}}{\bedp} P_Q \ \ \ \hbox{and} \ \ \ 
\tilde P_2 = -\frac{\bedpi{1}}{\bedp} P_Q
\end{equation}
and $\tilde Q_2 = -\tilde Q_1 = -Q$.
Plugging these into Eq.~\eqref{HsTOT} with $N=2$ leads to
\begin{equation}
\dHA^{(2,\text{s})} = \frac12\sqrt{-\bg} \bedpf \bcm^2 P_Q^2 - 
\frac{Q\bclap Q}{\bedpf\sqrt{-\bg} a^2}\\
+ \frac{\bcn^2 \bedpf}{3\bedp} \pMS P_Q, 
\end{equation}
where we have defined two different sound speeds
\begin{align}\label{eq:def:bcm}
\bcm^2 &\equiv \frac{\bcsi{1}^2\bedpi{2} +
  \bcsi{2}^2\bedpi{1}}{\bedp}, \\ \label{eq:def:bcn} \bcn^2 &\equiv
\bcsi{1}^2-\bcsi{2}^2,
\end{align}
and the conformal Laplacian is $\bclap \equiv a^2\blap$.

In order to add the adiabatic mode, we also define the operators
$\blapilapK \equiv \blap\bilapK$, and $\bclapK \equiv a^2\blapK$. The
three operators $\bclap$, $\blapilapK$ and $\bclapK$ are time
independent, i.e., the commutators $$[\lie_\bn, \bclap] = [\lie_\bn,
\bclapK] = [\lie_\bn, \blapilapK] = 0 $$ vanish when acting on any
tensor field. In terms of these, the final Lagrangian reads

\begin{equation}\label{ls3}
\dLA^{(2,\text{s})} = \pMS\MS^\prime + \ipS\iS^\prime - \dHA^{(2,\text{s})},
\end{equation}
with Hamiltonian
\begin{equation}
\begin{split}
\dHA^{(2,\text{s})} = &
  \pMS\frac{1}{2\om_\MS}\pMS + \ipS\frac{1}{2\om_S}\ipS\\ &+
    \MS\frac{\om_\MS\ofreq_\MS^2}{2}\MS +
    \iS\frac{\om_S\ofreq_S^2}{2}\iS \\ &+
    \frac{\bcn^2}{\bcs^2\bcm^2}\frac{\pMS\ipS}{\om_\MS\om_S{}
      \lapse{}H\blapilapK},
\end{split}
\label{hs3}
\end{equation}
where we are using the Hubble function notation $\bEX = 3H$ and
\begin{align}
\om_\MS &\equiv \frac{a^3\bedp}{\lapse\bcs^2\blapilapK H^2}, \qquad
\ofreq_\MS^2 \equiv -\frac{\lapse^2\bcs^2\bclap}{a^2}, \label{hs3:mf1}
\\ 
\om_S &\equiv \frac{1}{\lapse{}a^3\bcm^2\bedpf}, \qquad\, \qquad 
\ofreq_S^2 \equiv -\frac{\lapse^2\bcm^2\bclap}{a^2}.
\label{hs3:mf2}
\end{align}
We have also performed a change in the time variable such that $\dd{}t
= \lapse\dd\tau$, where $t$ is cosmic time, and $\lapse$ is an
arbitrary lapse function. Examining the Lagrangian in Eq.~\eqref{ls3},
one notes that the first two terms will simply change as, for
example, $$\dd{}t\dot{\MS} \rightarrow \dd\tau\MS^\prime,$$ where a
prime represents a derivative with respect to $\tau$, and the third term
will be multiplied by the lapse function, i.e., the Hamiltonian part of
the action is multiplied by $\lapse$. We have introduced the relevant
$\lapse$ factors directly on the definitions of the masses and
frequencies above.

To conclude this part, one sees, on the canonical form of the
Hamiltonian \eqref{hs3}, that two uncoupled fluids in an FLRW universe
end up being coupled through their interactions with the gravitational
field, and the coupling term has a definite form: it is a
momentum-momentum coupling whose amplitude is given by both sound
velocities.

\section{Quantization of Many Fields with Time-Dependent Quadratic
	Hamiltonians} \label{sec:Q}

In this section we will present the main ingredients to perform the
quantization of physical systems such as that presented in the previous
section. For details about single component quantization, we refer the
reader to Refs.~\cite{Vitenti2015} and Refs~\cite{Birrell1982, Wald1994,
Winitzki2005, Parker2009, Parker2012} for textbook treatments.

\subsection{The general approach}
\label{sec:genapp}

Let us suppose a physical system having $N$ different degrees of freedom
$$(\mf_1,\;\mf_2,\;\dots,\;\mf_N),$$
the fields $\mf_i$ describing, for instance, the cosmological perturbations
discussed above.
We define a solution in the phase space as the vector
$$
\phv_a = (\mf_1,\;\dots,\;\mf_N,\;\Pi_{\mf_1},\;\dots,\Pi_{\mf_N}),
$$
where $\Pi_{\mf_i}$ are the momenta canonically conjugate to $\mf_i$. 
If the Hamiltonian is quadratic in the fields, we  can write it in 
the form
\begin{equation}
\gH(\phv) = \frac12 \phv_a\gH^{ab}\phv_b,
\label{Hamil}
\end{equation}
where we introduced the symmetric Hamiltonian tensor $\gH^{ab}$. In what
follows, we shall also make use of the symplectic forms given by
\begin{align}
\SM_{ab} &\equiv
\ci\left( \begin{array}{cc} 0 & \mathbb{1}_{N\times N}
  \\ -\mathbb{1}_{N\times N} & 0 \end{array} \right),\\
\SM^{ab} &\equiv
\ci\left( \begin{array}{cc} 0 & \mathbb{1}_{N\times N}
  \\ -\mathbb{1}_{N\times N} & 0 \end{array} \right),
\end{align}
with $\mathbb{S}_{ac} \mathbb{S}^{cb} = \delta_a{}^b$. The phase space
indices are raised and lowered using the symplectic matrix, i.e.,
\begin{equation}
\phv^b \equiv \SM^{ab}\phv_b, \qquad \phv^{b*} \equiv \phv_a^*\SM^{ab}.
\end{equation}
The Hamiltonian equations for the system \eqref{Hamil} above, usually
written in the form
$$
\dot\mf_i = \frac{\partial\gH}{\partial\Pi_{\mf_i}} \ \ \ \ \hbox{and} \ \ \ \
\dot\Pi_{\mf_i}=- \frac{\partial\gH}{\partial\mf_i},
$$
then take the compact form
\begin{equation}
\ci\lie_\n\chi_a = \SM_{ab}\gH^{bc}\chi_c,
\label{LieSol}
\end{equation}
where $\n^\mu$ is the vector field which defines the foliation over
which the Hamiltonian is built.

The product of two solutions $\chi$ and $\varpi$, both assumed to
satisfy \eqref{LieSol}, is defined as,
\begin{equation}
\SM\Prod{\chi}{\varpi} \equiv \int_\ST\dd^3x \chi_a\varpi_b\SM^{ab},
\end{equation}
and it is conserved, i.e.,
\begin{equation}
\ci\lie_\n \SM\Prod{\chi}{\varpi} = 0,
\end{equation}
by virtue of the Hamilton equation \eqref{LieSol}, the antisymmetry of
$\mathbb{S}$ and the symmetry of $\mathcal{H}$.

Let us now quantize the theory through canonical quantization rules, and
consequently promote the fields to Hermitian operators. We will denote
the field operators by hats over the corresponding classical field:
$\Omf$ and $\Opmf$ will then be the field operators related to the
classical variables $\mf$ and $\pmf$, respectively. The canonical
commutation relations which one then imposes to quantize the system can
be regrouped in a single equation by means of the symplectic form $\SM$,
namely
\begin{equation}
\Comm{\Ophv_a(\bm{x})}{\Ophv_b(\bm{y})} = \SM_{ab}\dirac{3}{\bm{x}-\bm{y}}
\end{equation}
(recall we work in units where $\hbar = 1$).

We define the product of operators as
\begin{equation}
\Prod{\Ophv}{\hat{\gamma}} \equiv
\SM\Prod{\hat{\chi}^\dagger}{\hat{\gamma}},
\label{defprod}
\end{equation}
with $X^\dagger$ the Hermitian conjugate of $X$; this definition also
implies
\begin{equation}\label{DefProdScal}
\left(\vartheta,\upsilon\right) = \int_\Sigma \dd^3x\,\vartheta_a^* 
\left(\bm{x}\right) \mathbb{S}^{ab}\upsilon_b \left(\bm{x}\right)
\end{equation}
for ordinary phase space vectors $\vartheta_a$ and $\upsilon_a$: taking
the operators $\upsilon\hat{I}$ and $\vartheta\hat{I}$, with $\hat{I}$
is the identity operator, we can show that
\begin{equation}\label{eq:comm:prod}
\Comm{\Prod{\upsilon\hat{I}}{\Ophv}}{\Prod{\vartheta\hat{I}}{\Ophv}}
\!= \!\!\int\limits_\ST\dd^3x\vartheta_a^*(\bm{x})\upsilon_b^*(\bm{x})\SM^{ab}\hat{I}
= \Prod{\vartheta}{\upsilon^*}\hat{I}.
\end{equation}
In what follows, we now assume any function to be associated naturally
to an operator though the identity, and thus drop the $\hat{I}$.

Finally, we define the harmonic functions $\HF{\HFi}$ of the Laplace
operator,
\begin{equation}\label{eq:laplace}
\bscd^2\HF{\HFi} = -\EV{\HFi}^2\HF{\HFi},\quad \int_\ST\dd^3x
\HF{\HFi}\HF{\bm{p}} = \dirac{3}{\HFi-\bm{p}},
\end{equation}
where the index $\HFi$ represents all the indices necessary to label the
eigenvalues, and $\dirac{3}{\HFi - \bm{p}}$ represents the appropriate
Dirac and Kronecker deltas. We have also chosen to work with real
eigenfunctions of the Laplacian for simplicity (see~\cite{Vitenti2015}
for a discussion about this point). It is convenient to define a phase
space basis on $\HFi$-space (denoted with label $\HFi$), i.e.,
\begin{equation}\label{eq:def:basis}
\BF{\HFi,a}^i \equiv \BT^i{}_a (\HFi)\HF{\HFi},
\end{equation}
where we have $N$ complex phase space vectors $\BT^i{}_a (\HFi)$ or $2N$
real ones, the number of real solutions matching the initial conditions
degrees of freedom: $N$ fields and $N$ momenta. These solutions depend
only on time ($\bscd_\mu\BT^i{}_a = 0$) and the scale $\HFi$. These
vectors represent a complete set of solutions for the field equations
and can thus be viewed as analogous to tetrads. They therefore carry a
phase space index $a$ as well as a solution index $i$; there are as many
solutions as there are degrees of freedom, i.e., $N$ complex or $2N$
real solutions.

Given the definition \eqref{DefProdScal}, and the closure property
\eqref{eq:laplace}, the product of two basis vectors will be given by
\begin{equation}
\Prod{\BF{\HFi}^i}{\BF{\bm{p}}^j} =
\BT^{i}{}_{a}^*(\HFi)\BT^{j}{}_{b}(\bm{p})\SM^{ab}\dirac{3}{\HFi-\bm{p}},
\end{equation}
by virtue of Eq.~\eqref{eq:laplace}, so that, in order to have a
normalized basis, we must impose that
\begin{equation}
\begin{split}
\BT^i{}_a^*\BT^j{}_b\SM^{ab} = \delta^{ij}.
\end{split}
\label{orthonormal}
\end{equation}
For these orthonormal solutions, the natural ``metric'' is simply the
Kronecker delta $\delta_{ij}$, which we will use to raise and lower
solution indices, i.e.,
\begin{equation}
\BT_{ia} \equiv \delta_{ij}\BT^j{}_{a}.
\end{equation}

Given the orthonormal basis satisfying
$\Prod{\BF{\HFi}^i}{\BF{\bm{p}}^j} = \delta^{ij}\dirac{3}{\HFi-\bm{p}}$,
we can define the annihilation operators associated with this basis from
the field operator $\Ophv$ as
\begin{equation}
\AO_{\HFi}^i \equiv \Prod{\BF{\HFi}^i}{\hat{\chi}}.
\end{equation}
It follows directly from this definition and the properties above that
\begin{equation}
\Comm{\AO_{\HFi}^i}{\AO_{\bm{p}}^{j\dagger}} = 
\delta^{ij}\dirac{3}{\HFi-\bm{p}}.
\end{equation}
We also want to impose $\Comm{\AO_{\HFi}^i}{\AO_{\bm{p}}^{j}} = 0 =
\Comm{\AO_{\HFi}^{i\dagger}}{\AO_{\bm{p}}^{j\dagger}}$: this is
equivalent to demanding that $\Prod{\BF{\HFi}^{i*}}{\BF{\bm{p}}^j} = 0$.
For our choice of basis functions in the form given in
Eq.~\eqref{eq:def:basis}, one can see that the full set of conditions
reduce to
\begin{equation}
\begin{split}
\BT^{i}{}_a^*\BT^{ja} = \delta^{ij}, \qquad
\BT^{i}{}_a\BT^{ja} = 0.
\end{split}
\label{def:v:cond}
\end{equation}
A closer look at the definition of $\AO_{\HFi}^i$ shows that
\begin{equation}
\AO_{\HFi}^i = \BT^{ia*}\int_\ST\dd^3x
\HF{\HFi}(\bm{x})\hat{\chi}_a(\bm{x}) = \BT^i{}_a^*(\HFi)\tilde{\chi}^a(\HFi),
\end{equation}
where we have defined the transformed field $\tilde\chi_a\left(
\HFi\right)$ of the field operator as
\begin{equation}
\tilde{\chi}_a(\HFi) \equiv \int_\ST\dd^3x
\HF{\HFi}(\bm{x})\hat{\chi}_a(\bm{x}),
\end{equation}
whose commutator reads
\begin{equation}
\Comm{\OFphv_a(\HFi)}{\OFphv_b(\bm{p})} = \SM_{ab}\dirac{3}{\HFi-\bm{p}},
\end{equation}
and in terms of which one can write down the associated creation
operator through
\begin{equation}
\AO_{\HFi}^{i\dagger} = -\BT^i{}_a(\HFi)\tilde{\chi}^a(\HFi).
\end{equation}
The transformed field $\tilde\chi_a\left( \HFi\right)$ reduces to the
usual Fourier transform when $\HF{\bm{k}} \propto \ex^{\ci \bm{k}\cdot
	\bm{x}}$. This completes the quantum operator description of our
Hamiltonian system.


\subsection{Vacuum evolution}
\label{sec:vevol}

Let us assume that we define a vacuum at the instant $t_0$ using a given
basis $\BTi^i{}_{a}$ satisfying the conditions given in
Eq.~\eqref{def:v:cond}, i.e., $\BT^i{}_a(t_0) = \BTi^i{}_a$. At a later
time $t$, we can decompose the vector $\BT^i{}_a(t)$ in terms of
$\BTi^i{}_a$ as
\begin{equation}
\BT^i{}_a(t) = \alpha^i{}_j(t)\BTi^j{}_a +
\beta^i{}_j(t)\BTi^{j}{}_a^*,
\end{equation}
where the functions
\begin{equation}\label{eq:def:alpha:beta}
\alpha^i{}_j(t) \equiv \BT^i{}_a(t)\BTi_j{}^{a*} \quad \hbox{and}
\quad \beta^i{}_j(t) \equiv \BT^i{}_a(t)\BTi_j{}^a
\end{equation}
satisfy $\alpha^i{}_j(t_0) = \delta^i{}_j$ and $\beta^i{}_j(t_0) = 0$.
Consequently, the annihilation and creation operators at $t$ can be
written in terms of the corresponding operators at $t_0$ as
\begin{equation}
\begin{split}
\AO_{\bm{k}}^i(t) &= \alpha^i{}_j^*(t)\AO^j_{\bm{k}}(t_0) -
\beta^i{}_j^*(t)\AO^{j\dagger}_{\bm{k}}(t_0), \\ \AO_{\bm{k}}^{i\dagger}(t) &=
\alpha^i{}_j(t)\AO^{j\dagger}_{\bm{k}}(t_0) -
\beta^i{}_j(t)\AO^{j}_{\bm{k}}(t_0).
\end{split}
\end{equation}

It is clear from the equations above that if $\beta^i{}_j(t)$ vanishes,
or equivalently if $\Comm{\AO_{\bm{k}}^i(t)}{\AO_{\bm{p}}^j(t_0)} = 0$,
then both sets define the same vacuum. Writing the vacuum at a generic
time $t$ as $\ket{0_t}$, such that $\AO^i_{\bm{k}}(t)\ket{0_t} = 0$,
then the average number of particles at time $t$ given by the number
operator $N_{\bm{k}}^i(t) \equiv
\AO^{i\dagger}_{\bm{k}}(t)\AO^{i}_{\bm{k}}(t)$ which are present in the
initial vacuum state $\ket{0_{t_0}}$ reads
\begin{equation}\label{eq:unit:evol}
\braketOP{0_{t_0}}{N_{\bm{k}}^i(t)}{0_{t_0}} =
\dirac{3}{\bm{0}}\int\dd^3k\sum_j\left\vert\beta^i{}_j(t)\right\vert^2,
\end{equation}
where the factor $\dirac{3}{\bm{0}}\propto \mathcal{V}$ appears because
we are calculating the number of particles in an infinite spatial
section, with infinite volume $\mathcal{V}\to\infty$.
Reference~\cite{Vitenti2015} provides a more detailed discussion about this
infrared divergence in this context; the factor $\dirac{3}{\bm{0}}$ is
the same as that appearing in usual QFT in which it is accounted for by
considering a finite volume and taking the infinite volume limit
at the very last step \cite{Peskin:1995ev}.

As reviewed in Ref.~\cite{Vitenti2015}, in order for the quantum
evolution to be unitary, the number operator average given by
Eq.~\eqref{eq:unit:evol} must not diverge in the UV limit. 
Hamiltonians such as \eqref{hs3}, containing interaction terms mixing
momenta, can be problematic, because momenta exhibit a wavelength
dependence $\propto \EV{\HFi}^{1/2}$ initially, which can induce an
UV divergence when such interaction terms become relevant for
the dynamical evolution. In what follows, we present a formalism
permitting to eliminate such terms and subsequently express the result
in terms of Action-Angle (AA) variables, which will turn out to be more
suitable for understanding the frequency behavior of the relevant
degrees of freedom.

Consider a general Hamiltonian of the form $H=\frac12
\chi_aH^{ab}\chi_b$ with the bilinear Hamiltonian tensor given by
\begin{equation}
H^{ab} = \left( \begin{array}{cc} \POT & 0\\ 0 & \KIN\end{array} \right),
\label{HVT}
\end{equation}
where $\POT$ and $\KIN$ are arbitrary $N\times N$ matrices. Our starting
Hamiltonian tensor~\eqref{HVT} assumes no mixing between coordinates and
momenta. The Hamiltonian consisting of the sum of Eqs~\eqref{eq:HA:adia}
and \eqref{HsTOT} is of the required form \eqref{HVT}.
Although not the most general bilinear form one may think of, it
encompasses most of the cases relevant to cosmology, including the quite
general situations discussed in \cite{Langlois:2008qf}, and therefore a
thorough understanding of its properties provides an extremely useful
first step. We can express its mixed form as
\begin{equation}\label{HVTud}
\gH_a^{\ b} \equiv \SM_{ac}\gH^{cb} = \left(
\begin{array}{cc} 0 & \ci \KIN\\ -\ci \POT & 0\end{array} \right),
\end{equation}
from which one can solve the eigenvalue problem\footnote{We assume
wavelengths small enough compared to any relevant length scale of the
system to ensure that this eigenvalue problem makes sense.},
namely
\begin{equation}
\gH_a{}^c \Upsilon^{j}_{\ c} = \ofreq_{j} \Upsilon^{j}_{\ a} \qquad \hbox{and} \qquad
\gH_a{}^c \Upsilon^{j*}_{\ c} = -\ofreq_{j} \Upsilon^{j*}_{\ a},
\label{eigen}
\end{equation}
with $j=1,\cdots, N$ denoting the eigenvalue number; unless otherwise
explicitly stated, there is no summation on the eigenvalue number $j$.
It is worth noting that $\gH_a{}^c$ is a self-adjoint operator with
respect to the product, i.e.,
\begin{equation}
\begin{split}
\Prod{\phv}{\gH\cdot\vartheta} & = \Prod{\gH\cdot\phv}{\vartheta}, 
\\ \left(\gH\cdot\phv\right)_a & \equiv \gH_a{}^b\phv_b.
\end{split}
\end{equation}
In this case, the eigenvalues are real and the basis can be
made orthonormal.

Equations~\eqref{eigen} show that there are $N$ eigenvectors with
eigenvalues $\nu^j$, their complex conjugate having eigenvalues
$-\nu^{j}$. These eigenvectors can be written explicitly as
(we assume $\KIN$ to be invertible, which is always the
case in our context)
\begin{equation}
\Upsilon^{j}{}_a \doteq \left(\frac{q^j}{\sqrt{2\ofreq_j}}, -\ci \sqrt{\frac{\ofreq_j}{2}} \KIN^{-1} q^j\right),
\end{equation}
where $q^j$ satisfies
\begin{equation}
\KIN\POT q^j = \nu_j^2q^j,
\label{eq:qj:EV}
\end{equation}
and we have chosen $\ofreq_j > 0$. Also, one should keep in mind that
$\KIN$ is an $N\times N$ matrix and each $q^j$ an $N$ vector in
configuration space, so $\Upsilon^j$ is a $2N$ vector in phase space.
Expressed explicitly in components, the $q^j$ can be chosen in order to
yield
\begin{equation}
\Upsilon^{i}{}_a^* \SM^{ab}\Upsilon^j{}_b = \delta^{ij}.
\end{equation}
It is now possible to construct a real basis for vectors as
\begin{equation}
Q^{ja} = \frac{\Upsilon^{ja}+\Upsilon^{ja*}}{\sqrt{2m_j\nu_j}} =
\left(\frac{1}{\sqrt{\om_j}}\KIN^{-1} q^j,0\right)
\label{eq:Qja}
\end{equation}
and
\begin{equation}
\Pi^{ja} = \ci \sqrt{\frac{m_j\nu_j}{2}}\left( \Upsilon^{ja}-\Upsilon^{ja*}\right) =
\left(0,\sqrt{\om_j} q^j\right),
\label{eq:Pija}
\end{equation}
in which we introduced arbitrary functions of time $m_j$; we will
later use the freedom to choose these functions at will to deduce
the meaningful variables for which the time evolution of the
quantum field operators can be made unitary.

It is possible to expand the matrix $\SM$ on either basis, namely
\begin{equation}
\SM^{ab} = \sum_{j}\left( \Upsilon^{ja} \Upsilon^{jb*}-\Upsilon^{ja*}\Upsilon^{jb}\right)
\end{equation}
or
\begin{equation}
\SM^{ab} = \ci\sum_{j}\left( Q^{ja} \Pi^{jb}-\Pi^{ja}Q^{jb}\right),
\end{equation}
and the Hamiltonian tensor as
\begin{equation}
H^{ab} = \sum_{j} \nu_j \left( \Upsilon^{ja} \Upsilon^{jb*}+\Upsilon^{ja*}\Upsilon^{jb}\right).
\end{equation}
Now, defining the canonical variables
\begin{equation}
Q^j \equiv \chi_a Q^{ja} \qquad \hbox{and} \qquad \Pi^j \equiv \chi_a \Pi^{ja},
\end{equation}
one finds that the Lagrangian
\begin{equation}
\mathcal{L} = \frac{\ci}{2} \chi_a\SM^{ab}\dot\chi_b - \frac12 \chi_aH^{ab}\chi_b
\end{equation}
reduces to
\begin{equation}
\label{L1}
\begin{split}
\mathcal{L} = & \sum_{j} \frac12 \left( \Pi^j\dot Q^j - \dot\Pi^j Q^j \right)\\
&-\left[ \sum_{i} \frac{\Pi^{i2}}{2\om_i} + \frac12 \om_i\nu_i^2 Q^{i2}
+\sum_{i,j} \Pi^i \mathcal{M}^{ij}Q^j\right],
\end{split}
\end{equation}
where the matrix $\mathcal{M}^{ij}$ is defined through the time evolution
of the canonical variable $Q^j$, namely
\begin{equation}
\dot Q^{ja} = \sum_{i} \mathcal{M}^{ij} Q^{ia}.
\label{eq:def:M}
\end{equation}
Similarly, one could define the matrix $\mathcal{N}^{ij}$ for the
time evolution of the momentum through
\begin{equation}
\dot \Pi^{ja} = \sum_{i} \mathcal{N}^{ij} \Pi^{ia}.
\end{equation}
Using that $Q^{ia}\Pi^{j}_{\ a} = \ci \delta^{ij}$ and its time derivative,
one however finds that $\mathcal{M}^{ij}+\mathcal{N}^{ij}=0$, so
one really only requires knowledge of the matrix $\mathcal{M}^{ij}$.

The Hamiltonian appearing in the second line of Eq.~\eqref{L1} does not
contain quadratic terms mixing the momenta. This is not enough
however: interacting terms of the form $Q^i Q^j$ are those whose
UV limit behavior is under control, since they always lead to an
$\EV{\HFi}^{-1/2}$ power as leading term initially. Cross-terms mixing
the momentum and coordinate of the same degree of freedom can
be problematic however, and we need to perform a
further canonical transformation in order to eliminate all symmetric
mixing terms involving momenta. The transformation
\begin{equation}
\left\{
\begin{array}{l}
\Pi^{i} = P^i + \sum_{j} \mathcal{R}^{ij} Q^{j}, \\
\bar{Q}^i = Q^i,
\end{array}
\right.
\end{equation}
which is canonical if and only if $\mathcal{R}^{ij}=\mathcal{R}^{ji}$,
permits to achieve the goals discussed above (from here on we drop the overbar
in the field variable $\bar{Q}^i$). The Hamiltonian then reads
\begin{align}
\label{H2}
\mathcal{H} &= \sum_{i,j} \left\{\frac{P^{i2}}{2\om_i} + P^i \mathcal{M}^{ij}Q^j + \frac{P^i \mathcal{R}^{ij}Q^j}{\om_i}\right.+\frac{Q^{i}Q^j}{2}\times\nonumber\\
&\times\left.\left[\om_i\nu_i^2 \delta^{ij} + \dot{\mathcal{R}}^{ij} + \sum_{l} \left(\frac{\mathcal{R}^{il}\mathcal{R}^{lj}}{m_l}
+2\mathcal{R}^{il}\mathcal{M}^{lj}\right)\right]\right\}.
\end{align}

The symmetry condition $\mathcal{R}^{ij}=\mathcal{R}^{ji}$
is highly nontrivial since it actually imposes additional constraints
unfortunately preventing a full removal of the unwanted terms. In Eq.~\eqref{H2}
for instance, it would be the term containing $\mathcal{M}^{ij}$, and
the latter matrix is not necessarily symmetric, and so cannot
be fully canceled by the following term containing $\mathcal{R}^{ij}$.
Hence, another step is required: defining 
\begin{equation}
\mathcal{B}^{ij} \equiv 2\frac{\om_i\ofreq_i}{\ofreq_i+\ofreq_j} 
\mathcal{M}^{ij} = \mathcal{B}^{(ij)} + \mathcal{B}^{[ij]},
\label{Bij}
\end{equation}
one can choose the canonical transformation by imposing 
\begin{equation}
\mathcal{R}^{ij} = -\mathcal{B}^{(ij)},
\label{CCRij}
\end{equation}
yielding
\begin{equation}
\mathcal{H} = \sum_{i}\!\left(\frac{P^{i2}}{2\om_i} +
\frac{\om_i\ofreq_i^2 Q^{i2}}{2}\right) + \sum_{i,j} 
\!\left(\frac{P^i\tau^{ij}}{2\om_i\ofreq_i} - \frac{Q^{i}\gamma^{ij}}{2}\right)Q^j,
\label{H3}
\end{equation}
where
\begin{align}
\gamma^{ij} &\equiv \sum_{l} \left[\frac{\mathcal{B}^{(il)}\mathcal{B}^{(lj)}\ofreq_j}{m_l\ofreq_l}
+\frac{\mathcal{B}^{(il)}
\mathcal{B}^{[lj]}(\ofreq_l+\ofreq_j)}{\om_l\ofreq_l}\right]
+ \dot{\mathcal{B}}^{(ij)}, \label{gamma} \\
\tau^{ij} &\equiv \left[\mathcal{B}^{[ij]}(\ofreq_i+\ofreq_j)-\mathcal{B}^{(ij)}(\ofreq_i-\ofreq_j)\right],
\label{tau}
\end{align}
with $\tau^{ji}=-\tau^{ij}$. The matrices $\gamma^{ij}$ and $\tau^{ij}$
are the effective couplings between the various fluids. In the
single-fluid case, one often defines an effective time-dependent
frequency $\bar{\nu}$, which could be generalized in the multiple fluid
case through
\begin{equation}
\label{nubarra}
\bar{\ofreq}_{i}^2 = \nu_{i}^2 - \frac{\gamma^{ii}}{\om_i}.
\end{equation}
In the usual single-fluid case, Eq.~\eqref{nubarra} becomes 
$$\bar{\ofreq} = \ofreq -\displaystyle\left(\frac{\mathcal{B}^2}{m^2} + 
\frac{\dot{\mathcal{B}}}{m}\right),$$
which reduces to the effective potential $z''/z$ of
Ref.~\cite{Mukhanov1992} when one chooses $m\to 1$. In the case at hand
however, there are two extra terms that have to be taken care of.

Our final form \eqref{H3} could have been directly obtained with a
direct canonical transformation from Eq.~\eqref{eq:HA:final} instead of
going through the steps leading to Eqs.~\eqref{eq:HA:adia} and
\eqref{HsTOT}; it turns out that the adiabatic and entropy split can be
useful to compare with other formalisms, since it makes apparent the
usual Mukhanov-Sasaki variable. Besides, it allows for an easy
subsequent diagonalization since it sets the Hamiltonian in the block
diagonal form \eqref{HVT}.

The single term mixing coordinates and momenta in the Hamiltonian
\eqref{H3}, $\tau^{ij}$, being antisymmetric, can only relate the
momentum of a given fluid to the field variable of all other fields but
itself. As we anticipated above and will see below, the influence of
such kind of terms on the dynamics keeps the degrees of freedom
wellbehaved in the UV limit.

The coupling $\gamma^{ij}$ enters through a symmetric term, and one can
therefore restrict attention to $\gamma^{(ij)}$. Usually, one
incorporates its diagonal part into the effective mass, as in
Eq.~\eqref{nubarra}, leaving only a pure ``self-coupling'' term. Here,
we shall keep this term as it appears in Eq.~\eqref{H3} to prove that
its presence does not change the fact that the UV behavior still leads
to convergent $\beta$ functions, hence rendering possible a complete
definition of a vacuum state. For this to be true, one needs to choose
wisely the otherwise arbitrary functions $m_i$ to remove any potentially
problematic $\EV{\HFi}-$behavior. This is what we can see explicitly by
going to the Action-Angle variables $I_i$ and $\theta_i $.

The AA variables are defined through
\begin{equation}
\label{AA}
Q_{i} = \sqrt{\frac{2I_i}{\om_i\ofreq_{i}}} \sin\theta_i 
\qquad  \hbox{and} \qquad  P_{i} = \sqrt{2I_i \om_i\ofreq_{i}} \cos\theta_i,
\end{equation}
the functions $I_i$ being the momenta associated to the coordinates
$\theta_i$. It is worth mentioning that at this point we used the
original frequency $\ofreq_i$ instead of $\bar{\ofreq}_i$ (as was used
in~\cite{Vitenti2015}) in the implicit definition of $I_i$ and
$\theta_i$ above. This choice will result in a slightly different
approximation, but an equivalent scheme. Its usefulness arises
from the simplicity of the coupling terms in Eq.~\eqref{H4} which would
be spoiled by factors of $\ofreq_i/\bar{\ofreq}_j$.

The canonical variables \eqref{AA} allow us to express the Hamiltonian in
the required form, namely
\begin{equation}
\begin{split}
\mathcal{H} =& \sum_{i} 
 I_i\left[\ofreq_i + \dot{\alpha}_i\sin(2\theta_i)\right]\\
&\!-\sum_{i,j}
\sqrt{I_i I_j}\left[\sin\theta_i \sin\theta_j \bar{\gamma}^{ij}+ \frac{\sin(\theta_i-\theta_j) \bar{\tau}^{ij}}{2}\right],
\end{split}
\label{H4}
\end{equation}
where we defined 
\begin{align*}
\bar{\gamma}^{ij} &\equiv \frac{\gamma^{ij}}{\sqrt{\om_i \om_j \ofreq_i \ofreq_j}},
\qquad 
\bar{\tau}^{ij} \equiv \frac{\tau^{ij}}{\sqrt{\om_i \om_j \ofreq_i \ofreq_j}}, \\
\alpha_i &\equiv \ln\sqrt{\om_i\ofreq_i},
\end{align*}
from which one derives the equations of motion as
\begin{equation}
\dot{\theta_i} = \frac{\partial\mathcal{H}}{\partial I_i} \qquad \text{and} \qquad
\dot{I_i} = - \frac{\partial\mathcal{H}}{\partial \theta_i}.
\label{AAeqs}
\end{equation}
That this form of the Hamiltonian does not lead to divergent behavior of
the $\beta$ functions as discussed below Eq.~\eqref{eq:unit:evol} has
been proven in Ref.~\cite{Vitenti2015} for the single component case.
It also works for the $N$-fluids case, as we show below.

\subsection{Ultraviolet Behavior}
\label{sec:condini}

Let us now discuss the UV expansion of the solutions of
the Hamiltonian \eqref{H3}. We want to show that a particular choice of
initial conditions renders the respective $\beta$ functions such
that the particle number densities remain finite at all times. Arguably,
this choice of initial conditions provides a natural and well-defined
vacuum which generalizes the single-field case.

Let us begin by working out the spectral dependence of the various terms
involved in the UV limit $\EV{\HFi} \to \infty$. We are dealing
with general Hamiltonians of the form~\eqref{HVT}, which we want to
compare with the specific case of Eq.~\eqref{hs3}. We find that the
kinetic matrix $\KIN$ depends on the Laplacian only through
$\blapilapK$, and this, in the UV limit, is simply the identity. This
means that the asymptotic behavior of the kinetic matrix is
$\lim_{\EV{\HFi}\to\infty}\KIN = \OO{\EV{\HFi}^0}$. It is also easy to
see that for a set of scalar fields with canonical kinetic terms this
matrix will not depend on the Laplacian. For these reasons, we will
focus our attention to the case where $\lim_{\EV{\HFi}\to\infty}\KIN =
\OO{\EV{\HFi}^0}$. It is worth emphasizing that such a behavior
encompasses a large class of kinetic terms. Similarly, the potential
matrix in Eq.~\eqref{hs3} leads to terms proportional to the Laplacian,
so that its expected behavior in the UV limit is $\OO{\EV{\HFi}^2}$.
Again the same spectral dependence is found in the case of a set of
scalar fields whose couplings do not depend on spatial  derivatives.
Hence, we will focus on the case where $\lim_{\EV{\HFi}\to\infty}\POT =
\OO{\EV{\HFi}^2}$.

Using Eqs.~\eqref{eq:Qja} and \eqref{eq:Pija}, we have that the normalization
condition $Q^{ia}\Pi^{j}_{\ a} = \ci \delta^{ij}$ translates into
\begin{equation}
q^j\KIN^{-1}q^j = 1.
\end{equation}
Assuming the Hamiltonian to be ghost-free, the matrix $\KIN$ must
be positive 
definite. In this case, the asymptotic behavior of the 
vector $q^j$ is $\lim_{\EV{\HFi}\to\infty}q^j = \OO{\EV{\HFi}^0}$. From 
these UV behaviors, we conclude from Eq.~\eqref{eq:qj:EV} that 
$$\lim_{\EV{\HFi}\to\infty}\ofreq_i = \OO{\EV{\HFi}}$$ 
and, from Eqs.~\eqref{eq:def:M}, \eqref{gamma} and \eqref{tau},
we get
\begin{align*}
\lim_{\EV{\HFi}\to\infty}\mathcal{M}^{ij} &= \OO{\EV{\HFi}^0}, \qquad 
\lim_{\EV{\HFi}\to\infty}\bar{\gamma}^{ij} = \OO{\EV{\HFi}^{-1}}, \\
\lim_{\EV{\HFi}\to\infty}\bar{\tau}^{ij} &= \OO{\EV{\HFi}^{0}}.
\end{align*}
Finally, we can make use of our freedom in choosing $m_i$,\footnote{See~\cite{Vitenti2015} 
for a detailed discussion about this point.}  to ensure that
\begin{equation*}
\lim_{\EV{\HFi}\to\infty}\dot{\alpha}_i = \OO{\EV{\HFi}^{-2}}.
\end{equation*}
Since we are using $\ofreq_i$ instead of $\bar{\ofreq}_i$, one can see
that for the fluid Hamiltonian
Eqs.~(\ref{hs3}--\ref{hs3:mf2}), the frequencies $ \ofreq_i^2 $
are built from a time dependent function times the Laplacian. In this
case, we can choose $\om_i$ to cancel out this time dependent function
and provides $\om_i\ofreq_i = \EV{\HFi}$, and, consequently
$\dot{\alpha}_i = 0$. In our asymptotic analysis, we are always
considering the worst possible behavior (largest power of $\EV{\HFi}$)
of the variables involved. This will be, in turn, enough to show
that particle production is well-behaved. Once we know the asymptotic
behavior of every term in the Hamiltonian tensor Eq.~\eqref{H4} in the
UV limit, we can work out the required initial conditions.

Since we construct our variables from the Hamiltonian tensor
eigenvectors, all terms in Eq.~\eqref{H4} only depend on time
derivatives of the original Hamiltonian tensor components $\gH^{ab}$.
This means that even if the original Hamiltonian contains strong
couplings between the fields, the coefficients in Eq.~\eqref{H4} will be
small as long as the Hamiltonian tensor itself varies adiabatically
compared with the mode evolution. Let us introduce an ``adiabatic
parameter'' $\epsilon$ in the time derivative, i.e., $\lie_{\bn} \to
\epsilon\lie_{\bn}$, such that every time derivative of $\gH^{ab}$
produces a factor of $\epsilon$. Consequently, the quantities
$\dot{\alpha}_i$ and $\bar{\tau}^{ij}$ are all at least first order in
$\epsilon$, and $\bar{\gamma}^{ij}$ is at least of second order
[Eq.~\eqref{gamma}]. One can observe that $\bar{\gamma}^{ij}$, which
plays the role of the potential in the single component case, behaves as
a second order adiabatic quantity, while the coupling $\bar{\tau}^{ij}$
is a first order adiabatic quantity introduced by our first canonical
transformation which is only present in the many component case.

The equations of motion \eqref{AAeqs} yield, for the angles,
\begin{equation}
\begin{split}
\dot{\theta}_i &= \nu_i + \dot{\alpha}_i \sin(2\theta_i) 
-\bar{\gamma}^{ii}\sin(\theta_i)^2 - \frac{\mathcal{F}_i}{\sqrt{I_i}},
\label{thetaeq}
\end{split}
\end{equation}
where
\begin{equation}
\mathcal{F}_i \equiv \sum_{j\not= i}\!\!
\sqrt{I_j}\left[\sin\theta_i \sin\theta_j \bar{\gamma}^{ij} + 
\frac{\sin(\theta_i-\theta_j) \bar{\tau}^{ij}}{2}\right]\! .
\end{equation}
The adiabatic invariants $I_i$ then satisfy
\begin{equation}
\begin{split}
\dot{I}_i = & \left[\sin\left(2\theta_i\right) \bar{\gamma}^{ii}
-2\dot{\alpha}_i\cos\left(2\theta_i\right) \right]I_i  \\
&\! + \sum_{j\not= i} \sqrt{I_i I_j}\left[ 2\cos\left( \theta_i\right) 
\sin \left(\theta_j\right) \bar{\gamma}^{ij}
+\cos\left( \theta_i - \theta_j\right) \bar{\tau}^{ij} \right]\! .
\end{split}
\label{Ii}
\end{equation}
This set of equations potentially induces a technical difficulty. We
need a complete set of linearly independent solutions to construct all
necessary annihilation and creation operators [Eq.~\eqref{def:v:cond}].
Since the coupling terms are all at least of order $\epsilon$, then at
zeroth order, the adiabatic invariants are all constant
[Eq.~\eqref{Ii}]. It is tempting to choose each solution such that only
the $\ell^\text{th}$ adiabatic invariant is not vanishing at this order.
This choice produces a set of operators such that when setting
$\epsilon\to0$, we obtain that each creation/annihilation only affects
the $\ell^\text{th}$ field variable. In short, for the $\ell^\text{th}$
solution, all other adiabatic invariants $i\not=\ell$ are at least of
order $\epsilon$. The problem appears when we plug our solutions for
$I_i$ back into the equations of motion for the angles. In this
case the presence of the factor $I_i^{-1/2}$ in Eq.~\eqref{thetaeq}
turns these terms into zeroth order terms too. As a consequence, these
equations for the angles no longer decouple at this lowest order. We
will see below that this zeroth order term is actually signaling that
the subsidiary modes $i\not=\ell$ frequencies are indeed modified at
zeroth order.

We can circumvent this problem by introducing a new set of variables
representing the adiabatic invariant ``square-root'', namely
the complex fields $A_{\ell{}i}$ satisfying
\begin{equation}
A_{\ell{}i}A_{\ell{}i}^* = I_{\ell{}i},
\end{equation}
keeping in mind that these complex variables are not to be understood as
complexified solutions for the field, but merely a convenient form to
express the adiabatic expansion, i.e., $A_{\ell{}i} \equiv
\sqrt{I_{\ell{}i}}\ex^{\ci\theta_{\ell{}i}}$. We also introduced the
first index in the variables above to label the solution, i.e., the
first index specifies the solution and the second the phase space
component. This scheme will be maintained from here on. The equations of
motion for these variables are
\begin{equation}
\begin{split}
\dot{A}_{\ell{}i} = & \ci\left(\ofreq_i-\frac{\bar{\gamma}^{ii}}{2}\right)A_{\ell{}i} + \left(\frac{\ci\bar{\gamma}^{ii}}{2}- \dot{\alpha}_i\right) A_{\ell{}i}^* \\
&+ \frac{1}{2}\sum_{j\not=i}\left[\left(\bar{\tau}^{ij}-\ci\bar{\gamma}^{ij}\right)A_{\ell{}j} + \ci\bar{\gamma}^{ij}A_{\ell{}j}^*\right].
\end{split}
\label{dotAi}
\end{equation}
The effective frequency appearing in the first term of the equation
above can easily be identified as the first order expansion of the
frequency defined in Eq.~\eqref{nubarra}, i.e.,
$$
\bar{\ofreq}_i = \nu_i - \frac12 \bar{\gamma}^{ii} + \OO{\EV{\HFi}^{-2}}.
$$
Our choice of using the frequency $\ofreq_i$ instead of $\bar{\ofreq}_i$
makes the solutions differ at the order $\OO{\EV{\HFi}^{-2}}$. This is a
convenient choice for our purpose since it simplifies the coupling term
$\bar{\tau}^{ij}$ while not spoiling the convergence requirements.

Defining the angle variable
\begin{equation}
\sigma_i = \int_{t_0}^t\dd{}t\left(\nu_i - \frac12 \bar{\gamma}^{ii} \right),
\end{equation}
we rewrite the equations of motion as
\begin{equation}
\begin{split}
\dot{B}_{\ell{}i} = & \left(\frac{\ci\bar{\gamma}^{ii}}{2}- \dot{\alpha}_i\right)\ex^{-2\ci\sigma_i} B_{\ell{}i}^* \\
& + \frac12 \ex^{-\ci\sigma_i} \sum_{j\not=i}\left[\left(\bar{\tau}^{ij}-\ci\bar{\gamma}^{ij}\right)\ex^{\ci\sigma_j}B_{\ell{}j} + \ci\bar{\gamma}^{ij}\ex^{-\ci\sigma_j}B_{\ell{}j}^*\right],
\end{split}
\end{equation}
where $B_{\ell{}i} \equiv \ex^{-\ci\sigma_i}A_{\ell{}i}$. We can now write the adiabatic expansion as 
\begin{equation}
B_{\ell{}i} = \sum_{n=0}^\infty B^{(n)}_{\ell{}i},
\end{equation}
where $B^{(n)}_{\ell{}i}$ is of order $\OO{\epsilon^n}$. At zeroth order, the 
equation of motion is seen to reduce to
\begin{equation}
\dot{B}^{(0)}_{\ell{}i} = 0,
\end{equation}
whose solutions are constants, i.e., each mode oscillates with its
natural frequency through $\sigma_i$ at first order in $\epsilon$. As
discussed above, we may choose $B^{(0)}_{\ell{}i}(t_0) = 0$ for all
$i\not=\ell$ to obtain the $\ell^\text{th}$ solution and $
B^{(0)}_{\ell{}\ell}(t_0) = a_\ell$, for an arbitrary complex constant
$a_\ell$. For this reason, we need to go to the next order to obtain the
spectral dependency of the subsidiary modes $i\not=\ell$.

The equations at first order can be split into an equation for the phase
space component $\ell$,\footnote{In the following expressions we are
	keeping also second order adiabatic terms together with the first order
	ones for convenience.}
\begin{equation}
\begin{split}
&\dot{B}^{(1)}_{\ell{}\ell} = \left(\frac{\ci}{2} \bar{\gamma}^{\ell\ell}- \dot{\alpha}_\ell\right)
\ex^{-2\ci\sigma_\ell} a_\ell^{*},
\end{split}
\label{B1ll}
\end{equation}
and another for $i\not=\ell$
\begin{equation}
\begin{split}
&\dot{B}^{(1)}_{\ell{}i} = \frac12 \ex^{-\ci\sigma_i} \left[\left(\bar{\tau}^{i\ell}-\ci\bar{\gamma}^{i\ell}\right)
\ex^{\ci\sigma_\ell}a_\ell + \ci\bar{\gamma}^{i\ell}
\ex^{-\ci\sigma_\ell}a_\ell^{*}\right].
\end{split}
\label{B1li}
\end{equation}

Equation~\eqref{B1ll} can be solved by quadrature, i.e.,
\begin{equation}
\begin{split}
B^{(1)}_{\ell{}\ell} = B^{(1)}_{\ell{}\ell}(t_0) +  a_\ell^{*}\int_{t_0}^t\!\dd{}t\left(\frac{\ci}{2} \bar{\gamma}^{\ell\ell}- \dot{\alpha}_\ell\right)\ex^{-2\ci\sigma_\ell},
\end{split}
\end{equation}
which, once integrated by parts, leads to
\begin{equation}
B^{(1)}_{\ell{}\ell} = \ci{}a_\ell^{*}\left(\frac{\ci}{2}\bar{\gamma}^{\ell\ell} - \dot{\alpha}_\ell\right)\frac{\ex^{-2\ci\sigma_\ell}}{2\dot{\sigma}_\ell}.
\label{B1l}
\end{equation}
The other terms are proportional to time derivatives of the coupling
variables, and consequently are of higher order in the adiabatic
expansion. We can also use the freedom in choosing
$B^{(1)}_{\ell{}\ell}(t_0)$ to cancel out the contribution of the
integration by parts at $t_0$. Such a choice is justified since it is
the only one for which $B^{(1)}_{\ell{}\ell}$ goes to zero when
evaluated at a time at which the coupling goes to zero.
Equation~\eqref{B1l} shows that the first adiabatic correction to the
principal phase space variable $\ell$ is of order $\OO{\EV{\HFi}^{-2}}$
in the UV limit. 

A similar calculation for $i\not=\ell$ [Eq.~\eqref{B1li}] results in
\begin{equation}
\begin{split}
&B^{(1)}_{\ell{}i} = \frac12 \ex^{-\ci\sigma_i} \left( \frac{\ci\bar{\tau}^{i\ell} + \bar{\gamma}^{i\ell}}{\dot{\sigma}_i-\dot{\sigma}_\ell}\ex^{\ci\sigma_\ell}a_\ell - \frac{\bar{\gamma}^{i\ell}\ex^{-\ci\sigma_\ell}}{\dot{\sigma}_i +
\dot{\sigma}_\ell}a_\ell^{*}\right).
\end{split}
\end{equation}
Evaluating the UV limit, we obtain that $B^{(1)}_{\ell{}i}$ is of order
$\OO{\EV{\HFi}^{-1}}$. Finally, moving back to the original $A_i$,
we get
\begin{align*}
A_{\ell{}\ell} &= a_\ell\ex^{\ci\sigma_i} + \ci{}a_l^{*}\left(\frac{\ci}{2} \bar{\gamma}^{\ell\ell}- \dot{\alpha}_\ell\right)\frac{\ex^{-\ci\sigma_\ell}}{2\dot{\sigma}_\ell} + \OO{\epsilon^2}, \\
A_{\ell{}i} &= \frac{1}{2}\left(\frac{\ci\bar{\tau}^{i\ell}+\bar{\gamma}^{i\ell}}{\dot{\sigma}_i-
\dot{\sigma}_\ell}\ex^{\ci\sigma_\ell}a_\ell -
\frac{\bar{\gamma}^{i\ell}\ex^{-\ci\sigma_\ell}}{\dot{\sigma}_i +
\dot{\sigma}_\ell}a_\ell^{*}\right) + \OO{\epsilon^2}.
\end{align*}
Our choice of initial conditions makes the subsidiary modes oscillate
with the same frequency as the principal mode $\ell$. Moreover, the mode
$A_{\ell{}i}$ oscillates with both positive and negative frequencies
$\dot\sigma_\ell$. This behavior explains why the equation of motion for
the angle $\theta_{\ell{}i}$ is modified at zeroth order.

The equations above provide $N$ real solutions for our system of
equations. To form a complete set, we need $2N$ real solutions which we
use to build the $N$ complex phase space vectors $\BT^i{}_{a}$ presented
in~\eqref{eq:def:basis}. Given two real set of solutions
$(Q^\text{Re}_{\ell i}, \, P^\text{Re}_{\ell i})$ and
$(Q^\text{Im}_{\ell i}, \, P^\text{Im}_{\ell i})$,\footnote{These
variables should not be confused with the Hamiltonian tensor
eigenvectors given through Eqs.~\eqref{eq:Qja} and \eqref{eq:Pija}.} we
can write a complex set through
\begin{align*}
\BT^\ell{}_i = \frac{1}{2}\left[Q^\text{Re}_{\ell i}\left(I^\text{Re}_{\ell{}i}, \theta^\text{Re}_{\ell{}i}\right) - 
\ci Q^\text{Im}_{\ell i}\left(I^\text{Re}_{\ell{}i}, \theta^\text{Re}_{\ell{}i}\right)\right], \\
\BT^\ell{}_{i+N}  = \frac{1}{2}\left[P^\text{Re}_{\ell i}\left(I^\text{Re}_{\ell{}i}, \theta^\text{Re}_{\ell{}i}\right) - 
\ci P^\text{Im}_{\ell i}\left(I^\text{Im}_{\ell{}i}, \theta^\text{Im}_{\ell{}i}\right)\right].
\end{align*}
Each solution has its own adiabatic invariant and angle, i.e.,
$(I^\text{Re}_i, \, \theta^\text{Re}_i)$ and $(I^\text{Im}_i, \,
\theta^\text{Im}_i)$, respectively, for the solutions playing the roles
of the real and imaginary part. Note that, for the real part, we
use a different parametrization than that given in Eq.~\eqref{AA},
shifting the angle by $\pi/2$, as discussed in~\cite{Vitenti2015}, i.e.,
\begin{align}
Q^\text{Re}_{\ell{}i} &=
\frac{\sqrt{2I^\text{Re}_{\ell{}i}}}{\ex^{\alpha_i}}
\cos\theta^\text{Re}_{\ell{}i}, \quad P^\text{Re}_{\ell{}i} =
-\ex^{\alpha_i}\sqrt{2I^\text{Re}_{\ell{}i}}
\sin\theta^\text{Re}_{\ell{}i}, \label{AA2}\\ Q^\text{Im}_{\ell{}i} &=
\frac{\sqrt{2I^\text{Im}_{\ell{}i}}}{\ex^{\alpha_i}}
\sin\theta^\text{Im}_{\ell{}i}, \quad P^\text{Im}_{\ell{}i} =
\ex^{\alpha_i}\sqrt{2I^\text{Im}_{\ell{}i}}
\cos\theta^\text{Im}_{\ell{}i}.
\label{AA3}
\end{align}
In the limit $\epsilon \to 0$ the adiabatic invariants are constant and
the angles evolve solely with their respective frequency $\ofreq_i$. In
that case, if we set the same initial conditions for the angles, they will
remain equal, i.e., $\theta^\text{Re}_{\ell{}i} = \theta^\text{Im}_{\ell{}i}$
and, since the adiabatic invariants are constant, we can choose $$
I^\text{Re}_{\ell{}i} = I^\text{Im}_{\ell{}i} = \delta_{\ell{}i},$$
yielding \begin{equation}
\BT^\ell{}_i = \frac{\ex^{-\alpha_i-\ci\theta^\text{Re}_{\ell{}i}}}{\sqrt{2}}\delta_{\ell{}i}, \quad
\BT^\ell{}_{i+N} = -\ci\frac{\ex^{\alpha_i-\ci\theta^\text{Re}_{\ell{}i}}}{\sqrt{2}}\delta_{\ell{}i}.
\label{init:cond:0}
\end{equation}
In this case, in the solution $\ell$ only the $\ell^\text{th}$ variable
and its momentum [$(\ell+N)^\text{th}$ component] are different from
zero, and this set clearly satisfies Eq.~\eqref{def:v:cond}. This
amounts to showing that our choice of parametrization [Eqs.~\eqref{AA2}
and \eqref{AA3}] with identical initial conditions for the real and
imaginary parts results in the usual positive frequency solution for the
decoupled system.

It is a simple matter to obtain the shifted version of the equations of
motion using the new variables: setting 
$$ 
I^\text{Re}_{\ell{}i} = C_{\ell{}i}^* C_{\ell{}i}, \qquad
I^\text{Im}_{\ell{}i} = A_{\ell{}i}^* A_{\ell{}i},
$$ the choice $ C_{\ell{}i} = A_i\ex^{\ci\pi/2}$ yields
a modified version of Eq.~\eqref{dotAi}, namely
\begin{equation}
\begin{split}
\dot{C}_{\ell{}i} = & \ci\left(\ofreq_i-\frac12
\bar{\gamma}^{ii}\right)C_{\ell{}i} -
\left(\frac{\ci}{2}\bar{\gamma}^{ii}- \dot{\alpha}_i\right)
C_{\ell{}i}^* \\ &+ \frac12 \sum_{j\not=i}\left[\left(\bar{\tau}^{ij}-
\ci\bar{\gamma}^{ij}\right)C_{\ell{}j} -
\ci\bar{\gamma}^{ij}C_{\ell{}j}^*\right].
\end{split}
\label{dotCi}
\end{equation}
The only difference is the sign in front of the terms where the complex
conjugate of the field appears. As was done for $A_{\ell{}i}$, a similar
calculation can be performed, yielding
\begin{align*}
C_{\ell{}\ell} &= c_\ell\ex^{\ci\sigma_\ell} -
\ci{}c_\ell^{*}\left(\frac{\ci}{2}\bar{\gamma}^{ll} -
\dot{\alpha}_\ell\right)\frac{\ex^{-\ci\sigma_\ell}}{2\dot{\sigma}_\ell}
+ \OO{\epsilon^2}, \\ 
C_{\ell{}i} &= \frac12 \left( \frac{\ci\bar{\tau}^{i\ell}
	+ \bar{\gamma}^{i\ell}}{\dot{\sigma}_i -
	\dot{\sigma}_\ell}\ex^{\ci\sigma_l}c_\ell +
\frac{\bar{\gamma}^{i\ell}\ex^{-\ci\sigma_\ell}}{\dot{\sigma}_i +
	\dot{\sigma}_\ell}c_\ell^{*}\right) + \OO{\epsilon^2},
\end{align*}
where $c_\ell$ is an arbitrary complex constant.

From the approximate solutions $ C_{\ell{}i}$, we can extract the real
part of the solution and from $ A_i $ the imaginary. Setting $c_\ell =
a_\ell$, we obtain that $A_{\ell{}\ell}$ and $C_{\ell{}\ell}$ coincide
at zeroth order, and that their difference is of order $
\OO{\EV{\HFi}^{-2}} $. The same will be true for their module and
argument, i.e.,
\begin{align*}
\theta_{\ell \ell}^{\text{Re}} &= \theta_{\ell\ell}^\text{Im} + \OO{\EV{\HFi}^{-2}}, \\
I_{\ell\ell}^\text{Re} &= I_{\ell\ell}^\text{Im} + \OO{\EV{\HFi}^{-2}}.
\end{align*}
Consequently, for the principal mode we have
\begin{align}
\BT^\ell{}_\ell  &= \ex^{-\alpha_\ell}\left[\frac{\vert a_\ell\vert}{\sqrt{2}}
\ex^{-\ci\sigma_\ell} + \OO{\EV{\HFi}^{-2}}\right], \label{BTll}\\
\BT^\ell{}_{\ell+N}  &= \ex^{\alpha_\ell}\left[-\ci\frac{\vert a_\ell\vert}{\sqrt{2}}
\ex^{-\ci\sigma_\ell} + \OO{\EV{\HFi}^{-2}}\right],
\label{BTllN}
\end{align}
where we chose $ a_\ell $ real for simplicity. Note also that $a_\ell$
is a zeroth order quantity since it must be chosen in order to have
normalized states. Now, the subsidiary modes $i\not=\ell$ are given by
\begin{align}
\BT^\ell{}_i  &= \ex^{-\alpha_i}\left[ \ci\frac{\left\vert\bar{\tau}^{i\ell}
	a_l\right\vert\ex^{-\ci\sigma_\ell}}{\sqrt{2}\left\vert\dot\sigma_i -
	\dot\sigma_\ell\right\vert}  +
\OO{\EV{\HFi}^{-2}}\right], \label{BTli}\\ \BT^\ell{}_{i+N} &=
\ex^{\alpha_i}\left[\frac{\left\vert\bar{\tau}^{i\ell}
	a_l\right\vert\ex^{-\ci\sigma_\ell}}{\sqrt{2}\left\vert\dot\sigma_i -
	\dot\sigma_\ell\right\vert} +
\OO{\EV{\HFi}^{-2}}\right].
\label{BTliN}
\end{align}
Here we kept only the term involving $ \bar{\tau}^{il} $, this 
term being proportional to $ \OO{\EV{\HFi}^0} $, whereas the 
coupling $ \bar{\gamma}^{il} $ is of order $\OO{\EV{\HFi}^{-1}}$,
therefore contributing only in the next order in the UV expansion.

It is worth emphasizing that the explicit form of the coupling
containing the coefficient $\bar{\tau}^{ij}$ in the Hamiltonian
[Eq.~\eqref{H4}] is what generates the subsidiary solution term of order
$\OO{\EV{\HFi}^{-1}}$. However, since this term does not couple
$A_{\ell{}i}$ with $A_{\ell{}j}^*$ [Eq.~\eqref{dotAi}], it does not
change when we shift the equations by $\pi/2$ to obtain the equations of
motion for $C_{\ell{}i}$. As a result, it creates an identical term in
both the real and imaginary adiabatic invariants and angles at first
order. In contrast, $\bar{\gamma}^{ij}$ does not appear in such special
form and, therefore, it couples $A_{\ell{}i}$ with $A_{\ell{}j}^*$
generating different factors [of order $ \OO{\EV{\HFi}^{-2}} $] in the
real and imaginary solutions. This behavior of the coupling
$\bar{\tau}^{ij}$ is essential to obtain a well defined particle
production, and it is a consequence of our choice of $R^{ij}$ given by
Eqs.~\eqref{Bij} and \eqref{CCRij}.

In short, the principal mode $ \BT^\ell{}_\ell $ and its momentum $
\BT^\ell{}_{\ell+N} $ are composed of a zeroth order [$ \OO{\EV{\HFi}^0}
$] term oscillating with the same complex exponential plus second order
[$\OO{\EV{\HFi}^{-2}}$] corrections. The subsidiary modes $i\not=\ell$
and their momenta are composed of a first order term
[$\OO{\EV{\HFi}^{-1}}$] oscillating with the same complex exponential
plus second order terms [$\OO{\EV{\HFi}^{-2}}$].

Given our choice of initial conditions for the adiabatic corrections
discussed below Eq.~\eqref{B1l}, the subsidiary modes go to zero when
evaluated at a time where the couplings also go to zero. This results in
the solutions approaching Eq.~\eqref{init:cond:0} in this time
limit.\footnote{If the couplings never vanish, it is always possible to
choose the initial conditions for the adiabatic corrections such that
the subsidiary modes are exactly zero at a given time. In this case, it
is reasonable to choose the initial time as that at which the couplings
are minimized.} Consequently, these solutions satisfy the conditions
present in Eq.~\eqref{def:v:cond} in the limit, choosing $a_\ell = 1$.
Thus, since the product is constant, it will satisfy the conditions for
any time $t$. In other words, these solutions define a set of
creation/annihilation operators with the required commutation relations
where modes are statistically independent of each other.

Before moving forward, it is important to understand why quantization of
Hamiltonian \eqref{eq:HA:adia} $+$ \eqref{HsTOT} in the original
variables, i.e., the adiabatic plus entropy modes variable split, would
not be well defined. Take, for example, the two fluid Hamiltonian of
Eq.~\eqref{hs3}. The coupling between modes appears as a product of the
momenta times a time dependent function of order $\OO{\EV{\HFi}^0}$.
When written in terms of the adiabatic invariants, each momentum will
contribute a factor of $\OO{\EV{\HFi}^{1/2}}$ [see Eq.~\eqref{AA}]. For
this reason, the final coupling between adiabatic invariants will be of
order $\OO{\EV{\HFi}^1}$. If the original coupling between momenta is
small in a given time limit, one can perform the expansion in powers of
the coupling as we have done for the adiabatic parameter. On the other
hand, in this case, the subsidiary term is proportional to the coupling
times a factor of order $\OO{\EV{\HFi}^1}$. Now, applying the limit
where the coupling goes to zero, we obtain the vacuum commonly used in
the literature. However, since this term is also proportional to
$\OO{\EV{\HFi}^1}$, if we first perform the limit $ \EV{\HFi}\to\infty
$, we obtain infinity, violating the assumption that the coupling is
small. Put in another way, the decoupling limit and the UV limit do not
commute for this choice. Given small but finite couplings, there will
always be an infinite range of UV modes where the small coupling
approximation is not valid, so that such a vacuum definition is not UV
complete. In contrast, our choice of variables generates couplings
between the adiabatic invariants that are at most of order
$\OO{\EV{\HFi}^0}$. Therefore, in our case, the adiabatic and UV limits
commute and a well-defined vacuum can be calculated for all UV modes.

The original form \eqref{eq:HA:final} provides a well-defined
next-to-leading order approximation, yet it leads to a non-unitary
evolution. The adiabatic/entropy split, leading naturally to the
Hamiltonian \eqref{eq:HA:adia} $+$ \eqref{HsTOT}, would seem to be even
more problematic at first sight since it not only leads to a non-unitary
evolution but actually has an ill-defined next-to-leading order
approximation for the base functions.

In practice however, it turns out to be more convenient since it is in a
block-diagonal form and thus easier to transform into the final,
well-defined choice of variables with Hamiltonian \eqref{H3}, for which
the evolution is unitary. Indeed, the  coupling terms in
\eqref{eq:HA:final} between the field variable and momenta do not
satisfy the required symmetries. Therefore, they mix $A_{\ell{}i}$ and
$A_{\ell{}i}^*$ at order $\OO{\EV{\HFi}^{-1}}$ such that the subsidiary
terms of this order no longer have equal real and imaginary adiabatic
invariants and angles. Consequently, the presence of such terms
generates a $\OO{\EV{\HFi}^{-1}}$ term in the particle creation
operator, which would thus not converge.

With the variables leading to \eqref{H3}, the leading approximation is
given by Eqs.~(\ref{BTll}--\ref{BTliN}), so that every basis function
for the $\ell$ mode oscillates with the same frequency $\sigma_\ell$. In
\eqref{eq:HA:final}, the effective coupling $\bar{\tau}^{ij}$ is not
antisymmetric, and therefore the above-mentioned subsidiary terms would
depend on both $\ex^{-\ci\sigma_l}$ and $\ex^{\ci\sigma_l}$, thereby
mixing positive and negative frequencies at leading order. This
indicates particle creation at $\OO{\EV{\HFi}^{-1}}$ order, and hence,
the non unitary evolution advertised above. Ideally, the
adiabatic/entropy intermediary mode expansion could be avoided by means
of a direct transformation from \eqref{eq:HA:final} to \eqref{H3}; in
general however, going through this step provided a straightforward
method to transform any block-diagonal Hamiltonian into the required
form.

We are now in a position to evaluate whether the $\beta^{ij}(t)$
calculated through Eq.~\eqref{eq:def:alpha:beta} present the required
convergent UV behavior, thereby providing acceptable initial conditions
for a vacuum state. From Eq.~\eqref{eq:def:alpha:beta}, one obtains
$\beta^{ij}(t) = \BT^i{}_a(t) \BTi^j{}^a$, or, more explicitly,
\begin{align}
\beta^{ij} &= \ci\sum_r\left(\BT^i{}_r\BTi^j{}_{r+N}-
\BT^{i}{}_{r+N} \BTi^j_{r}\right).
\label{betat2}
\end{align}
Each term of this expansion of $ \beta^{ij} $ depends on the factor of
$\ex^{\alpha_i-\alpha_i(t_0)}$. This is exactly the problematic factor
present in the single component quantization, discussed thoroughly in
Ref.~\cite{Vitenti2015}. There it was shown that our choice of $\om_i$
is enough to make this factor behave as
$\ex^{\alpha_i-\alpha_i(t_0)} = 1 + \OO{\EV{\HFi}^{-2}}$. These factors
lead to convergent integrals and therefore can be safely ignored.
Finally, for every mode, the leading term is symmetric when changing the
time dependence, i.e.,
$$ 
\BT^i{}_r\BTi^j{}_{r+N} = \BTi^i{}_r\BT^j{}_{r+N} + 
\OO{\EV{\HFi}^{-2}},
$$
and consequently, every term in the sum defining $ \beta^{ij} $ will be
at least of order $ \OO{\EV{\HFi}^{-2}} $. This amounts to showing that,
for this choice of initial conditions and variables, the particle number
density will be finite at all times and, therefore, that the time
evolution of the quantum field operators will be unitary. 

We stress at this point that it is not the leading order that would
produce the divergent contribution for the particle creation, but the
next-to-leading order: the leading order can be made to cancel out by an
appropriate choice of the arbitrary mass functions appearing in Eqs.
\eqref{eq:Qja} and \eqref{eq:Pija}, but then the problem reappears at
the following order for which no such arbitrarily adjustable functions
is available. We also observe that is not necessary that all adiabatic
corrections be of order $\OO{\EV{\HFi}^{-2}}$: if it was indeed the
case, the term $\bar{\tau}^{ij}$ would have spoiled the convergence.
What is however necessary is that any coupling mixing $A_{\ell{}i}$ and
$A_{\ell{}i}^*$ (consequently also $C_{\ell{}i}$ and $C_{\ell{}i}^*$)
frequencies must be at least of order $\OO{\EV{\HFi}^{-2}}$. We have
thus achieved our goal to obtain a consistent set of initial conditions
leading to an unambiguous vacuum.

\section{Conclusion}

Setting consistent and natural initial conditions for cosmological
perturbations crucially depends on the model one is investigating. In
inflationary models, very short wavelengths, i.e., those much smaller
than the almost-constant Hubble scale, can always be seen as effectively
evolving in a Minkowski universe, and can thus be given a natural vacuum
initial state (or any other state one wishes \cite{Martin:1999fa}).
Nonetheless, even in this case, not all canonical variables lead to a
unitary evolution of the quantum field, as was elucidated in
\cite{Vitenti2015}. In more complicated situations involving many
fluids, things are worse, as the presence of subdominant modes creates
new possibilities for divergences in the particle production. The
present work goes one step further in completing the analysis by
clarifying the situation when many components are present.

In the first part of this work, we managed to obtain the Hamiltonian for
a set of fluids coupled only through gravitation. This Hamiltonian was
obtained without ever using the background equations of motion. For this
reason, one can use it even when the background is also quantized
(see~\cite{Peter2007, Peter2008} for a single fluid example) or in other
settings where the background does not satisfy the classical equations
of motion; another example is when the background is described by an
averaged metric. One should also stress that the ambiguity in defining
the gauge invariant perturbations discussed in~\cite{Vitenti2013} is
also present in the many fluids case, so that our formalism is necessary
to define the gauge invariant perturbation whenever one uses a
nonclassical background.

The second part consists in establishing the general procedure for many
component quantization. We restricted attention to cases where the
original Hamiltonian tensor can be written in the form~\eqref{HVT} and
the spectral behavior of its many components described in
Sec.~\ref{sec:condini}. This restriction is however quite mild, as this
encompasses most systems found in the literature. We showed that the
usual procedure through which one determines the vacuum by expanding the
solutions in terms of the original couplings is potentially ill defined.
For example, if the original coupling terms appear as products of the
different component momenta in the Hamiltonian, then the basis determination
is incomplete in the UV limit, i.e., there always exists a value of
$\EV{\HFi}$ beyond which the coupling becomes strong and the whole
approximation invalid. Consequently, the UV limit does not commute with
the small coupling limit. This shows that determining the basis
functions for quantization of multiple component systems using only the
adiabatic expansion (or slow-roll approximation in a multifield
inflation scenario) can be misleading since the corrections can have a
divergent UV limit. In addition, we proved that with our choice of
variables, the time evolution can be implemented by a unitary operator.
This result extends the literature~\cite{Torre2002,Corichi2006,
Cortez2007, Corichi2007,BarberoG2008,Vergel2008,Cortez2011,
Gomar2012,Cortez2012,Fernandez-Mendez2012,Cortez2013, Cortez2015,
Vitenti2015} by introducing for the first time a set of special
canonical variables for a multiple component system coupled by quadratic
terms in the Hamiltonian for which the time evolution is unitary. All
these results were accomplished using adiabatic invariants, which proved
essential to understand the UV asymptotic behavior of the solutions.

Our formalism can in particular be applied to the special case where
only two fluids are present, in which case one needs not specify a
privileged fluid with respect to which the other would be defined.
Instead, one expands in the usual adiabatic and entropy modes, for which
one then finds very natural vacuum (or else) initial conditions, where
the quantum evolution is unitary. One can apply straightforwardly our
result to specific cases involving, for instance, two fluids in a
contracting and bouncing universe. That will be the subject of a
subsequent work \cite{PPVPrep}.

\section*{ACKNOWLEDGMENTS}

We would like to thank CNPq of Brazil and ILP for financial support.
S.D.P.Vitenti acknowledges financial support from Capes, under the
program ``Ci\^{e}ncias sem Fronteiras'' (Grant No. 2649-13-6). We
also wish to thank S.~Renaux-Petel for enlightening discussions.

\bibliography{references}
\bibliographystyle{apsrev4-1}

\end{document}